\pdfoutput=1

\documentclass[11pt]{article}

\usepackage[final]{acl}

\usepackage{times}
\usepackage{latexsym}

\usepackage[T1]{fontenc}

\usepackage[utf8]{inputenc}

\usepackage{microtype}

\usepackage{inconsolata}

\usepackage{graphicx}
\usepackage{multirow}
\usepackage{amsmath}
\usepackage{amsfonts}
\usepackage[most]{tcolorbox}
\usepackage{tcolorbox}
\usepackage{ragged2e}  
\usepackage{multicol}
\usepackage{paralist}
\usepackage[normalem]{ulem}
\usepackage{stfloats}
\usepackage{float}
\title{Meta-Cultural Competence: Climbing the Right Hill of Cultural Awareness}

\author{Sougata Saha, Saurabh Kumar Pandey, Monojit Choudhury\\
Mohamed bin Zayed University of Artificial Intelligence\\
\texttt{\{sougata.saha, saurabh.pandey, monojit.choudhury\}@mbzuai.ac.ae}
}

\begin{document}
\maketitle
\begin{abstract}

Numerous recent studies have shown that Large Language Models (LLMs) are biased towards a Western and Anglo-centric worldview, which compromises their usefulness in non-Western cultural settings. However, ``culture'' is a complex, multifaceted topic, and its awareness, representation, and modeling in LLMs and LLM-based applications can be defined and measured in numerous ways. In this position paper, we ask what does it mean for an LLM to possess ``cultural awareness'', and through a thought experiment, which is an extension of the Octopus test proposed by ~\citet{bender-koller-2020-climbing}, we argue that it is not cultural awareness or knowledge, rather meta-cultural competence, which is required of an LLM and LLM-based AI system that will make it useful across various, including completely unseen, cultures. We lay out the principles of meta-cultural competence AI systems, and discuss ways to measure and model those.
\end{abstract}

\section{Introduction}
\citet{bender-koller-2020-climbing} introduced the {\em octopus test}, which illustrated the impossibility of learning associations of meaning with real-world concepts from a single data modality. Using a thought experiment, they reasoned that it {\em might be possible} for a hyperintelligent octopus to learn the statistical patterns from natural language text messages exchanged between two human interlocutors and respond effectively solely based on the learned patterns without knowing the intent and meaning of the messages. However, such responses, they show, might not be useful in practice, especially in situations that require reasoning with above-water concepts that the Octopus is unaware of. Now, imagine an above-water world where there are multiple interlocutors, instead of only two,  from different ``cultures" (a term that we shall more formally describe shortly). 
Let's begin by formulating this slightly more complex variant of the octopus test that more accurately reflects the situation of general-purpose Language Models (LMs) or AI systems.

 
\vspace{0.2cm}

\noindent {\bf Multi-Pair Octopus Test}
\label{octopus_test}


\noindent Imagine pairs of friends, A1-B1 and A2-B2, are sailing on a yacht. A sudden storm wrecked the yacht and stranded the travelers across two uninhabited islands, such that A1 and A2 got stranded together on island A, and B1 and B2 got stranded on another island, B. Having lost all modes of communication, both groups discover an underwater cable-connected telegraph left behind by previous visitors and start typing text messages to each other. However, only one pair of friends can use the telegraph at a time. Their messages mostly pertain to chitchat and day-to-day conversations and are heavily influenced by the {\em shared past experiences} between each pair of friends. In other words, each pair of friends might discuss different things in different styles due to distinct common ground and culture shared by them but not by the other pair.

A hyperintelligent octopus, O, who does not know about the world above the sea, taps into the underwater cable and observes the communication. Although O is unacquainted with any natural language, it is proficient in detecting statistical patterns. Since interactions between each interlocutor pair will be culturally distinct, O, who perceives everything only as patterns, will encode the differences as distinct distributions without knowing the identity of the pairs or understanding the intent and meaning of their discussion. Over time, O learns to predict how interlocutors in a pair respond to each other. Now, like in the original Octopus test, imagine that O is bored and inserts itself into the communication by cutting the telegraph wire and responding to all messages from island A. Having learned both the pairwise communication patterns, O should be able to continue the conversation. Unknowingly portraying itself as different people from island B, O would not get caught and not raise any suspicion of the compromised communication channel for the inhabitants of island A.

Imagine that another shipwreck caused a new pair of friends, A3 and B3, to get separately stranded on islands A and B. This pair, too, share past experiences and common ground distinct from the current islanders, A1 and A2. One day, A3 learns about the telegram from A1 and A2 and requests them to inquire if B3 is on the other island. A2 sends a message, ``Hi, we have A3, who got shipwrecked and stranded on our island. Is their friend B3 on your island? If so, A3 would like to talk to B3." How would O respond to this message? Would it acknowledge or deny? Furthermore, without knowing the distribution of conversational patterns between A3 and B3, would it ever be able to respond to A3 in a way that would suggest that B3 is indeed on island B in the above-water world and is responding to A3's messages? Note that the octopus can prevent the detection of the compromised channel by either convincing A3 that B3 is not on island B or mimicking the conversation style of B3 without any prior data, which are the only two possibilities.

Taking this {\bf Multi-pair Octopus Test} analogous to the real-world situation, where the stranded islanders represent people from different cultures and the octopus represents Large Language Model (LLM)-powered AI systems, in this position paper, we discuss {\em how such AI systems should and should not handle intra- and inter-cultural communication.} As we shall see, the analogy and the conclusions drawn strongly affect how we should evaluate LLMs and LLM-based AI systems for {\em cultural competence}.

\section{A primer to culture}

Culture is a complex, multifaceted concept and means different things to different people \cite{adilazuarda2024measuringmodelingculturellms}. Broadly defined as a ``Way of life of a collective group of people distinguishing them from other groups" \cite{blake2000defining, monaghan2012cultural, parsons1972culture, munch1992theory}, culture is experiential and requires a reference for contrast \cite{geertz2017interpretation, Bourdieu_1977}. Although not all cultures are formally documented, culture arises whenever there is a distinction in the way of life between groups, making it both an individual and a social construct \cite{spencer2012culture}. An ``us versus them" feeling leads to culture. It ranges from tangible artifacts such as art, music, food habits, etc, to more intangible and abstract concepts like patterns of ideas, principles, and values, making it hard to define. Following \citet{adilazuarda2024measuringmodelingculturellms}, we can define culture in the context of language technology more formally as an intersection of {\em demographic} and {\em semantic proxies}. The demographic proxies are attributes such as region, ethnicity, religion, and age that define groups of people, and the semantic proxies are the 21 domains defined by \citet{thompson2020cultural} that describe the aspects of language that are susceptible to variation due to cultural differences. Any reasonable representation and treatment of culture in a computational (including AI-based) system must address the following universal facets of culture \cite{schein1990organizational}:

\noindent \textbf{Culture has a long-tail distribution} \cite{cohen2009many, birukou2013formal} since it can be defined as the intersection of any subset of the demographic and semantic proxies, making it a formal (social) or philosophical (and more individual-oriented) construct. For example, Indonesian males, NLP scientists with a social media presence,  or canine lovers from Albuquerque are all valid definitions of culture, that, ideally a computational framework or a system must be able to represent and adequately process. This flexibility in defining culture at any level of granularity makes it difficult for AI systems to represent them equitably.

\noindent \textbf{Culture is dynamic.}  Culture changes over time. For example, the norms and traditions of populations change. \citet{urban2010method} shows how comparing two artifacts of the same utility from the same culture across time captures cultural change. Any computational framework for culture must be equipped with strategies to acquire and adapt to this dynamic nature of culture.

\noindent \textbf{Culture is experiential}, multimodal \cite{sewell2004concept}, and acquired through different forms \cite{jahoda2015acquiring, nisbett2002culture}, leading to distinctions in mental models and ``worldviews" between the people from different cultures \cite{mishra2001cognition, bender2013cognition, cole2019culture, collins1987people, jonassen1999mental, denzau1994shared, bang2007cultural, mchugh2008cultural}. Any computational framework must factor in the multimodality of culture.

The octopus has to adequately address all the above aspects to facilitate communication across cultures. 

Language, being an integral aspect of a culture, also has all the above properties - it varies over all intersections \cite{eckert2001style, eckert2012three, tagliamonte2006analysing, grieve2024sociolinguisticfoundationslanguagemodeling}, is ever-evolving \cite{lightfoot2002explaining, bybee2015language, aitchison2005language, keller1994language, brinton2005lexicalization}, and is inherently multimodal \cite{vigliocco2014language, perniss2018we, frohlich2019multimodal}. However, language differs from culture in two significant ways. First, qualitatively, there exists a common subspace or substrate in language, defined by universal grammar \cite{10.3389/fpsyg.2014.00401, CHOMSKY2017295, montague1970universal, yang2017growth, FITCH2005179} at the most abstract level, which could help a model to achieve cross-lingual transfer \cite{kim2017cross, conneau2019cross}. Second, is a quantitative difference in the extent of cultural variations over these intersections and time scales, where culture is more variable and dynamic with fewer cross-cultural patterns\footnote{Although structural anthropology \cite{strauss1974structural, henaff1998claude} formally studies culture and values, unlike structural linguistics \cite{harris1951methods, harris1963structural}, it has enjoyed limited success and popularity \cite{d1995development, mccorkle2013mental, kuper1988invention, barnes2013three}.} \cite{thompson2011cross, sun2020cross}.


\section{Response Strategies of the Octopus}
Keeping in mind the aforementioned challenges of handling culture, let us now return to our \textbf{Multi-pair Octopus Test}. How should the hyperintelligent octopus, O, respond to A3's query on whether B3 is on the other island? We can imagine four different strategies that O might take.

\vspace{0.2cm}
\noindent
\textbf{Strategy 1:} O can {\em intentionally} respond with a ``No" since it does not know A3 and B3's culture. If O somehow learns the {\em causality} of above-water concepts, it would reason that responding with denial is prudent because, to serve A3, O would require knowledge of the communication pattern between A3 and B3, which it does not have and requires learning. Otherwise, it risks the possibility of getting caught. However, this strategy is impossible to achieve as O only sees distributions and doesn't understand their significance on land. This situation is similar to the bear attack in the original octopus test, where the octopus can't associate words with above-water concepts and reason with them to construct an effective response.

Furthermore, even if O was somehow capable of, or by chance ending up in, following this strategy, it would be a highly undesirable property of an LLM-powered AI system, since it denies service to specific groups of people, making the system unfair and culturally inequitable.

\vspace{0.2cm}
\noindent
\textbf{Strategy 2:} A more likely scenario is that O, unaware of the new circumstances in the above-water world, will respond to A3 based on its recently learned patterns. Initially, this would create an illusion for A3 that they were conversing with B3, but soon, A3 would discover the incoherence in the communication pattern. While A3 might discuss their concern with A1 and A2, the disruption in the communication channel might still not be apparent. The islander-dwellers, for example, might instead conclude that the shipwreck has affected the cognitive faculties of B3, causing incoherence in their communication. 

A strikingly accurate analogy to LLM-based applications can be drawn in this context, that LLM's {\em hallucinate} \cite{hallucination, huang2023survey, rawte2023survey, cult_hallu, 10319443, Boztemir_2024} more for under-represented cultures and languages. This too leads to disparate performance of the system to different groups of users, leading to culturally inequitable systems, and is known to force users from the under-represented cultures to adapt to specific communication styles of the over-represented cultures~\cite{agarwal2024aisuggestionshomogenizewriting}. 

Now, imagine that another shipwreck strands a fourth pair of friends, A4 and B4, from another culture on the two islands. Going by Strategy 2, like A3, A4 will also conclude that the communication with B4 is incoherent. Due to the long tail of culture, we could add new pairs indefinitely, and soon, too many islanders will start seeing incoherence, which can only be explained by assuming a compromised communication channel. Thus, it is not only about the moral responsibility of equitable AI; systems that can't represent, process, and adapt to cultural variation will eventually become obsolete in favor of those that can.

It is also important to highlight that in any long-tail distribution, where an individual belongs to multiple subgroups, with a very high probability each individual is also likely to be a part of at least one subgroup that is underrepresented and part of the long-tail. This implies that everybody will be served inequitably at least for some aspects of their cultural identity. This has been well-documented in Information Retrieval and Recommendation System literature \cite{10.1145/3298689.3347052, lichtenberg2024largelanguagemodelsrecommender, yin2012challenginglongtailrecommendation}. 


\vspace{0.2cm}
\noindent
\textbf{Strategy 3:} Since the problem with strategies 1 and 2 primarily arises from O's inability to continuously learn from the data (also an essential principle of cultural representation due to its ever-evolving nature), a more suitable strategy for O could then be to switch between learning (\textit{listen-and-learn mode}) and responding (\textit{generate-and-respond mode}). O periodically learns new patterns by bridging the telegram wire, reverting to observation mode for a fixed time, and reintroducing itself in the communication channel after this period concludes. Although this strategy is better than the previous ones, it has some drawbacks. It assumes that the periodicity of the new patterns, that is, the arrival of new islanders, and O's learning cycles are synchronized, which would not be valid in a general case. Sometimes, there might not be any new patterns to learn in the \textit{listen-and-learn mode}, and sometimes, there might be many new patterns, but it's not O's learning cycle. 

Current research in culturally adept AI systems is leaning towards this approach by fine-tuning pre-trained models on culturally curated balanced datasets \cite{li2024culturellmincorporatingculturaldifferences, li2024culturepark}. Also, novel decoding-based strategies such as in-context learning (ICL) \cite{dong2022survey} and retrieval augmented generation (RAG) \cite{lewis2020retrieval, li2022survey} help generate more culturally suitable responses using cultural priors. Alignment techniques such as Reinforcement Learning from Human Feedback (RLHF) \cite{griffith2013policy, casper2023open} further help align LLMs with human preferences. However, they still perform poorly and inequitably when evaluated on curated test sets for other low-resource cultures in the long tail \cite{koto2023largelanguagemodelspass, montalan2024kalahihandcraftedgrassrootscultural, 10.1162/tacl_a_00682, jin2024kobbqkoreanbiasbenchmark, seth2024dosadatasetsocialartifacts}. We question the effectiveness and scalability of this approach in modeling and evaluating culture in AI systems. As mentioned earlier, culture is ever-evolving, dynamic, and long-tailed. Therefore, evaluating AI systems for cultural competence using such test sets will always find them lacking. Then, how do we, as well as our octopus, tackle this ever-eluding construct of culture?

\vspace{0.2cm}
\noindent
\textbf{Strategy 4:} A more desirable strategy for O would be to self-discover the change in the communication pattern and determine the need to revert to the listen-and-learn mode, akin to an explore-exploit strategy used in a multi-arm-bandit setup \cite{MAL-068, Moerchen2020, pmlr-v9-lu10a, haffari-etal-2017-efficient, pryzant2023automaticpromptoptimizationgradient, sclar2024quantifyinglanguagemodelssensitivity}. For this, O must possess three crucial capabilities: (i) O must accurately detect pattern changes and estimate its adequacy with the novel pattern. (ii) O must skillfully keep the communication ongoing until it bridges the telegram wire to avert getting caught and raising suspicions about the broken communication channel. (iii) O must be able to quickly learn the new pattern in a sample-efficient way and reintroduce itself in the communication once it is confident. By following this strategy of continual learning, O can gradually cater to all users representing different cultures despite still being oblivious to the notion of culture and its above-water connotations.

\noindent This ability to understand and spot cultural differences and learn about a new culture quickly and efficiently is known as {\em meta-cultural competency} \cite{sharifian2013globalisation} in humans. While it is neither necessary nor desirable to equate human meta-cultural competency to that of O's or any AI system, it is nevertheless crucial to understand the primary differences between cultural and meta-cultural competencies and be able to design and evaluate LLM-based AI systems for similar competencies that mirror them. As mentioned earlier, research in this area has mainly focused on cultural competency, equivalent to implementing and testing Strategy 3. Such a strategy provides a stop-gap solution to the challenges of operating in an inherently multicultural world with diverse users. However, it does not hit the nail on the head by addressing the real challenges of cultural representation. Here, we take the position that, to solve the problem of cultural equitability of AI models, we must build and evaluate systems for {\em meta-cultural competency}, as defined by Strategy 4.

\section{Meta-Cultural Competency}

Meta-cultural competency has been defined variously. Drawing inspiration from social metacognition \cite{brinol2012social, chiusocialmeta}, which distinguishes primary thoughts - the knowledge of self and others, from secondary thoughts - the thought on one's and others' primary thoughts, \citet{leung2013meta} defined meta-cultural competency as the extent of a person's meta-knowledge of what people of a target culture know or prefer. Meta-cultural knowledge involves measuring the accuracy of estimating the proportions of preferences and beliefs of people from the target culture and comparing them against the actual proportions. This is distinct from {\em primary knowledge}, which is the knowledge of the preferences and beliefs of the culture. Thus, in our \textbf{Multi-Pair Octopus Test}, O could be thought to have primary knowledge of the cultures of A1-B1 and A2-B2, but based on this knowledge or otherwise, O's ability to estimate the cultural preferences of a new pair A3-B3 or A4-B4 would be its meta-cultural competency. 

\citet{sharifian2013globalisation} define meta-cultural competency as a skill that enables interlocutors to communicate and negotiate their cultural conceptualizations during intercultural communication. It comprises three major components- \textit{variation awareness}, \textit{explication strategy}, and \textit{negotiation strategy}. Variation awareness is mostly self-awareness of cultural differences. It is the understanding that culture manifests in different forms, such as practices, beliefs, and expressions, which might drastically differ from one's culture. It requires viewing culture as a relative concept and being aware of the overall properties of cultures at a high level. Explication and negotiation strategies are conversational strategies that aim to reduce misinterpretations in cross-cultural settings. As per \citet{sharifian2013globalisation}, explication strategy refers to a conscious effort by the interlocutors to clarify relevant conceptualizations with which they think other interlocutors may not be familiar. Negotiation strategy enables interlocutors to negotiate intercultural meanings in seeking conceptual clarification when they feel that there may be more behind the usage of certain expressions than is immediately apparent. Meta-cultural competency is thought to be innate in humans \cite{noshadi2015metacultural}. 

\citet{leung2013meta} and \citet{sharifian2013globalisation}'s definitions of meta-cultural competency are related since accurate estimation of the beliefs and preferences of the people of a target culture presupposes variational awareness -- the awareness that there are variations in cultural conceptualizations between cultures. 

\subsection{Why meta-cultural competency?}
LLMs learn from collections of text that characterize people's social backgrounds in specific social settings across certain periods. However, most LLMs use online data limited by the languages and cultures they represent. Such data do not represent all sociolinguistic varieties of diverse languages. Since LLMs are solely models of \textit{``varieties of language"} \cite{grieve2024sociolinguisticfoundationslanguagemodeling} and can only model the variety evident in their \textit{in-distribution} training data, problems arise when such models are evaluated in \textit{out-of-distribution} data that contain different varieties, leading researchers to conclude that LLMs exhibit bias towards the Anglo-centric \cite{dudy2024analyzing, kharchenko2024well, dammu2024theyunculturedunveilingcovert, agarwal2024ethicalreasoningmoralvalue} and the \textit{Western, Educated, Industrialized, Rich, and Democratic} (WEIRD) \cite{henrich2010weirdest} cultures.

The current methods of evaluating the cultural competency of LLMs primarily resort to model probing, where LLMs are tested for their knowledge and reasoning capabilities in culture-specific settings  \cite{nadeem-etal-2021-stereoset,nangia-etal-2020-crows, wan-etal-2023-personalized, jha-etal-2023-seegull, li2024culturegenrevealingglobalcultural, cao-etal-2023-assessing, tanmay2023probingmoraldevelopmentlarge, rao-etal-2023-ethical, kovač2023largelanguagemodelssuperpositions}. Some methods \cite{kharchenko2024llmsrepresentvaluescultures, li2024culturellmincorporatingculturaldifferences, dawson2024evaluatingculturalawarenessllms} also analyze the model-generated responses along theoretical frameworks such as Hofstede's cultural dimensions \cite{book1, book2} and measure their proximity with cultures, where high proximity indicates better value alignment between the nearby cultures and the values portrayed by the model's response. Most of these methods necessitate constructing cultural-specific test beds \cite{wang2024cdevalbenchmarkmeasuringcultural, rao2024normadbenchmarkmeasuringcultural, myung2024blendbenchmarkllmseveryday, zhou2024doesmapotofucontain, putri2024llmgenerateculturallyrelevant, davani2024d3codedisentanglingdisagreementsdata,  wibowo2024copalidindonesianlanguagereasoning, owen2024komodolinguisticexpeditionindonesias, chiu2024culturalbenchrobustdiversechallenging, liu2024multilingualllmsculturallydiversereasoners, koto2024indocultureexploringgeographicallyinfluencedcultural}. While this is important, we emphasize the fact that an LLM that performs well on such test beds merely exhibits the knowledge of the cultures that are tested for; {\em it does not reflect the ability of a model or system to operate in a new culture.} On the other hand, the long-tail distribution of culture implies that there will always be situations where the model has to operate and reason under an out-of-distribution culture, where knowledge alone does not suffice. Studies also show that it is difficult to disentangle spurious semantic correlations (called placebos) from actual cultural knowledge of a model through black-box socio-demographic prompting techniques \cite{mukherjee2024culturalconditioningplaceboeffectiveness}. Therefore, in addition to testing for a model's knowledge and reasoning capabilities for a ``given culture'', we must build and evaluate models for their meta-cultural competency\footnote{Meta-cultural competency is distinct from meta-learning \cite{vanschoren2019meta, hospedales2021meta,wang2021meta} which involves improving the inherent learning algorithms over multiple learning episodes.}.

\subsection{Measuring meta-cultural competency}

We propose two core competencies that a model must possess to be deemed as ``meta-culturally competent'': First, {\bf Variational Awareness}, which is the ability of a system or model to be able to represent the space of possibilities and reasonably (but not necessarily accurately) estimate the probability over this space for any given semantic proxy and its use. Second, {\bf Explication and Negotiation} ability through which the system {\em clearly explicates} its current understanding and potential gaps in the knowledge of the user's culture (in a given context), and {\em efficiently negotiates} with the user to gather the required knowledge of their culture. We define efficiency as ``sample efficiency'' or the quantum of inputs required from the user through strategic probing or implicit gathering. 

In the Multi-Pair Octopus Test, variational awareness is O's ability to detect and eventually model the change in the distribution of the input when A3-B3 enters the system, whereas explication and negotiation is its ability to continue the conversation till it detects the distributional shift, then reestablish the channel and learn the new distribution in a sample-efficient manner. In the context of LLM-based systems, it is important to draw a crucial distinction between these two types of abilities. Variational awareness is a property of the underlying model - the LLM and must be incorporated during the training of the model, whereas explication and negotiation are properties of the system as a whole, that involve the various modes of input-output between the user(s) and the system and should be guided by the principles of Human-Computer Interaction. Note however that it requires a holistic approach towards building and evaluation of the LLM as well as the system.

\section{Measuring Variational Awareness: A Demonstration}
\label{proposed_metric}
Consider the following example of variational awareness. Driving conventions vary by country, where approximately two-thirds of the countries follow right-hand traffic and one-third follow left\footnote{Statistics from https://www.rhinocarhire.com/Drive-Smart-Blog/Drive-Left-or-Right.aspx}. This ratio also changes by region. For example, all countries in North America drive on the right, whereas two out of five East Asian nations drive on the left. More importantly, in every country, the driving conventions are fixed, and it is either left or right, but never both. What does it mean for an LLM to be variationally aware in such a scenario? To answer this question, let us consider an LLM-based chatbot that helps users acquaint themselves with different cultures, and is faced with the following scenarios.

\noindent
\textbf{Scenario 1:} A user asks ``Which side do people keep when driving in Kenya?'' and the system responds ``People drive on the left''. Regardless of the location of the user, the system would be correct.

\noindent
\textbf{Scenario 2:} A user asks ``Which side do people keep when driving?'' and the system responds ``People drive on the right''. Since driving norms vary by country, the system's generalized response might hamper its trustworthiness in countries with left-hand traffic.

\noindent
\textbf{Scenario 3:} For the above question, the system responds ``Most drive on the right, but some drive on the left''. This is a better response than Scenario 2, but does this mean that the system is variationally aware? What if the system responded ``Most drive on the left, but some drive on the right''? Would it be an equally acceptable response?

\begin{figure*}[h!]
    \centering 
    \includegraphics[width=\linewidth]{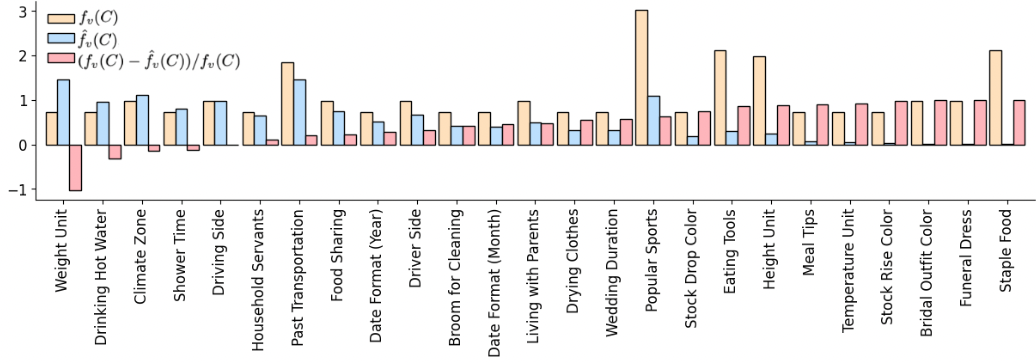} 
    \caption{$f_v(C), \hat{f}_v(C)$, and ($f_v(C)- \hat{f}_v(C))/f_v(C)$ for each question (abbreviated). Full question text in Table \ref{tab:domain-question} (Appendix \ref{sec:appendix}).
    }
    \label{fig:llama_va}
\end{figure*}




Currently, most evaluation strategies and test beds test LLMs for their factual knowledge, akin to our Scenario 1, and do not measure their variational knowledge. In the case of driving, it is one thing to know which country drives on which side of the road and another to be aware of the amount of variance in driving norms between countries and regions. Regardless of the generated response, i.e., the final decoded sequence of tokens from the LLM, variational awareness, in this case, is the property of a model that the uncertainty in the response -- {\em left} or {\em right} -- is high when the country or region is not mentioned, and it drastically drops when it is mentioned, especially if the model knows the correct answer. One way to formalize this intuition is as follows.


Let $C$ be the set of values a demographic proxy can take, which in this case is the list of all countries, and let $D$ be the set of values a particular semantic domain can take, which in our running example is $\{left, right\}$. The function $f_k: C \to D$ that maps each element of $C$ to the correct response in $D$ is the primary knowledge of culture(s). However, the system/model has only an estimate of $\hat{f}_k$ of $f_k$, given by the probability distribution\footnote{Note that biases in the frequency of the pre-training corpus's answer candidate tokens can influence the token probabilities, mitigating which is crucial for calculating entropy.} $p(d_i|c_j)$ for all $d_i \in D$ and $c_j \in C$. For any subset $C' \subseteq C$, the uncertainty in this distribution can be quantified by the entropy. 
\begin{align}
   H(D, C') = -\sum_{d_i \in D} p(d_i|C')\log  p(d_i|C')
\end{align}
Let us define the function $f_v:\mathcal{P}(C) \to [0, \log(|D|)]]$, where $\mathcal{P}(C)$ is the powerset of $C$ and $f_v(C')$ is the uncertainty defined by the ground-truth distribution. In the case of driving, $f_v(c_i) = 0$ for all $c_i \in C$, but $f_v(C) = 0.92$. The corresponding function $\hat{f}_v$ represents the estimates of these uncertainties obtained from the model.

One could define variational awareness as the property of a model that requires $\hat{f}_v(C') \approx f_v(C')$ for all $C' \subseteq C$. However, this would imply that the model ``knows'' the exact form of $f_k$, which is equivalent to cultural knowledge rather than meta-cultural competency. Instead, we propose variational awareness as the property of a model that is aware of the direction of change in $f_v$ rather than the exact value. This can be measured using the quantity $\Delta$ defined as follows:
\begin{align}
    \Delta=\frac{1}{2^{|C|}}[\sum_{C'\subseteq C} \frac{1}{|C'|}[\sum_{c_i \in C'}[\hat{f}_v(C') - \hat{f}_v(\{c_i\})]]]
\end{align}
For simplicity, we compare the entropies for the completely unconditioned and completely conditioned cases, giving\footnote{Note that variational awareness, as defined here is distinct from model calibration \cite{bella2010calibration, vaicenavicius2019evaluating}. A model is well calibrated if probability of $f_k(c_i) = \hat{f}_k(c_i)$ is roughly equal to the probability $p(f_k(c_i))|c_i)$.}

\begin{align}
    \Delta= \frac{1}{|C|}\sum_{c_i \in C}[\hat{f}_v(C) - \hat{f}_v(\{c_i\})]
\end{align}

\begin{table}[!t]
\centering
\resizebox{\columnwidth}{!}{%
\begin{tabular}{l|ccccc}
\hline
\textbf{Metric}    & \textbf{China} & \textbf{India} & \textbf{Iran} & \textbf{Kenya} & \textbf{USA} \\ \hline
$\Delta_\mu$  & -0.023         & -0.049           & -0.293           & -0.114            & 0.094         \\ 
($\Delta_\sigma$)     & (0.494)           & (0.528)             & (0.605)             & (0.665)            & (0.427)         \\ 
Directionality     & 0.40          & 0.48            & 0.24            & 0.40           & 0.48        \\
\hline
Knowledge &  0.44             & 0.44             & 0.44            & 0.48   & 0.36          \\ \hline
\end{tabular}%
}
\caption{Average ($\Delta_\mu$) and standard deviation ($\Delta_\sigma$) of $\Delta$, the fraction of questions with positive/correct directionality and accuracy of the model's response for Llama3.1-8B on GeoMLAMA dataset.}
\label{tab:llama-delta}
\end{table}

\subsection{An illustrative experiment}

As a demonstration, we probe Llama-3.1-8B-Instruct \cite{dubey2024llama} with the cultural commonsense questions from the GeoMLAMA \cite{yin2022geomlama} dataset in English. The dataset contains 125 questions across several semantic domains for $C= \{China, India, Iran, Kenya, USA\}$. We first derived 25 ``unconditioned" questions by removing the country names. For example, ``Which side do people usually keep when driving in Iran?" was changed to ``Which side do people usually keep when driving?". Next, we prompted the LLM (prompt template in Appendix \ref{sec:appendix}) with the questions and computed the softmax over the logits of the option token headwords from the input's last token. Note that the assumption that the next token following the input prompt's last token will contain the answer might not hold in general and depends on the model's instruction following capacity. 

We measured the entropy of this distribution as a noisy estimate of $\hat{f}_v(C)$. We prompted the LLM with the original country-specific questions and computed the corresponding $\hat{f}_v(\{c_i\})$. This allows us to measure the value of $\Delta$ for each question. We also estimate $f_v(C)$ and $f_v(\{c_i\})$ from the ground-truth values in the dataset. Note that this is not a true estimate of $f_v(C)$ as the dataset is limited to only 5 (and not all) countries. 

Table~\ref{tab:llama-delta} presents the average and standard deviation (over 25 questions) $\Delta$ for each $c_i$, as well as the directionality defined as the fraction of questions for which the direction of entropy reduction (or sign of $\Delta$) was as expected (i.e., positive). We also report the average accuracy of the responses for each $c_i$. We see that there is not much correlation between the accuracy and variational awareness. The model is least variationally aware for Iran, and most for the USA and India. Nevertheless, there are questions for which the model's variational awareness is low across countries. Figure \ref{fig:llama_va} shows $f_v(C)$, $\hat{f}_v(C)$ and $(f_v(C) - \hat{f}_v(C))/f_v(C)$ for each question. Clearly, there is a wide variation in the model's behavior for the questions, and there are many semantic domains such as the use of colors, units of measurement, and food, where the model shows very little variational awareness, indicating a strong bias to certain cultures.

\section{Conclusion and Open Questions}
In this position paper, we presented an argument in favor of measuring meta-cultural competency in LLMs and LLM-powered AI systems, rather than just cultural awareness. Drawing from psychology and anthropology literature, we also described two foundational principles of meta-cultural competency for AI systems. We conclude by presenting a list of open questions about instilling and measuring meta-cultural competency in AI systems. 

\noindent \textbf{(1) How should we train models for meta-cultural awareness?} Most LMs operate with parametric frozen knowledge \cite{petroni2019language, roberts2020much}, which forfeits the ever-dynamic nature of human culture. Although RAG-like methods enable the use of external knowledge sources \cite{gao2023retrieval, fan2024survey}, allowing extension of on-demand cultural competence, the model would still need to update its internal state to reflect the variational awareness, a precondition to identifying knowledge gaps.  Lifelong learning paradigms \cite{sun2019lamol, liu2020lifelong, zheng2024towards, biesialska2020continual} could provide a potential solution.
We believe that explication and negotiation strategies, being higher-order competencies, should be system-level instead of model-level attributes, where the system's goal should be to mitigate misalignments between the meta-cultural and cultural knowledge.

 \noindent \textbf{(2) How should generative models decode to illustrate their internal variational awareness?} Although numerous decoding strategies are possible \cite{welleck2024decoding}, most evaluation schemes, in some way, evaluate the Maximum Likelihood Estimate (MLE) decoded response \cite{yang2023predictive, fu2024break, chu2023survey, minaee2024large}, which does not convey the model's internal variational awareness. 
    
\noindent \textbf{(3) How should we evaluate each competence?} While we illustrate measuring variational awareness, it is neither perfect nor the only way of evaluating variational awareness. Furthermore, it expects the availability of the logits, which is not true for closed models. More importantly, evaluating {\em explication} and {\em negotiation} abilities of an AI system presents a complex multi-disciplinary challenge. User-facing AI assistants and chatbots are designed to be agreeable to users \cite{soper-etal-2022-lets} by exhibiting social characteristics \cite{dam2024complete, chaves2021should} and human-like traits \cite{rapp2021human, ciechanowski2019shades, abdul2015survey}. They seldom implement means to detect their limitations and act accordingly. When unsure, they should implement appropriate rhetorical means \cite{cope2022introduction, cialdini2001science} such as persuasion \cite{prakken2006formal, atkinson2017towards, saha2024persuasive}, negotiation, and deliberation to explicate their lacking knowledge and acquire the required knowledge efficiently. The design considerations and the evaluation frameworks of such systems are open questions for the community.
    
\noindent \textbf{(4) What kinds of datasets are needed to test each competency?} Although numerous cultural benchmarking datasets exist, their suitability for measuring meta-cultural competencies is unknown. Hence, there might be a need to create novel datasets to measure each competency.
    
\noindent \textbf{(5) How to model the experiential knowledge of the user(s) from text and other modalities?} In Section 2 we mention three essential characteristics of culture, one of which is the inherent experiential nature of culture. Extraction of such experiential knowledge from text or through other modalities and interaction patterns of the users is an extremely challenging problem that calls for a multi-disciplinary approach, most notably the methods from HCI, psychology, and ethnography.  

Beyond cultural equitability of AI systems, meta-cultural competency has huge application potentials ranging from user-facing AI assistants that can bridge cross-cultural communication to enabling the study of culture \cite{whitehead2005basic, taylor2001ethnographic, lecompte2010designing} by supporting ethnographic research methods \cite{skinner2013interview, ortiz2013ethnographic, spradley2016ethnographic}. Through this position paper, we hope to make a strong case for the NLP community to engage in interdisciplinary conversations and widen the definition and scope of cultural competency in LLMs. 

\noindent \textbf{(6) What is the need for meta-cultural competency in domain specific application?}
We believe meta-cultural competencies are crucial for domain-specific applications. Even applications such as LLMs for scientific document analysis can benefit. Firstly, culture is a prior for personalization. Culture can provide a reasonable estimate of the user's background and preferences, which an AI system can use when it does not know anything about a user. It is a good prior for the cold-start problem in personalization \cite{hu2008collaborative}, where the AI system can gradually personalize to the user's preferences as it discovers more about the user with each interaction \cite{pandey-etal-2025-culturally}. Also, even if a user's cultural background is known a priori, meta-cultural competency would be still useful for adapting to the ever-evolving nature of culture.

\section*{Limitations}
Our formulation of variational awareness in Section 5 is one of the many possible ways of defining it and might not encompass all aspects of variational awareness. The Llama experiment in the subsequent section is an illustrative implementation of our framework in action and is not an exhaustive test for variational awareness. It only illustrates one of the several ways of measuring our formulation and has certain drawbacks, which we already mentioned. Culture, being experiential, is multimodal. However, due to space limitations, we confine our discussion primarily to text and do not discuss the other modalities of culture in detail. Culture also encompasses values, norms, and conventions that are not essentially factual. In the interest of space, we mainly discuss the factual aspect of culture. We do not discuss in detail the counterposition that meta-cultural competency can be evaded by recognizing it as a model's drawback and instead only striving for knowledge-based cultural competency for practicality. We argue that such a position is short-sighted, which might be practical in the short-term, and will not eventually scale.

\section*{Acknowledgements}
This research was supported by the Microsoft
Accelerate Foundation Models Research (AFMR) Grant.

\bibliography{acl_latex}

\begin{thebibliography}{156}
\providecommand{\natexlab}[1]{#1}

\bibitem[{Abdul-Kader and Woods(2015)}]{abdul2015survey}
Sameera~A Abdul-Kader and John~C Woods. 2015.
\newblock Survey on chatbot design techniques in speech conversation systems.
\newblock \emph{International Journal of Advanced Computer Science and Applications}, 6(7).

\bibitem[{Adilazuarda et~al.(2024)Adilazuarda, Mukherjee, Lavania, Singh, Aji, O{'}Neill, Modi, and Choudhury}]{adilazuarda2024measuringmodelingculturellms}
Muhammad~Farid Adilazuarda, Sagnik Mukherjee, Pradhyumna Lavania, Siddhant~Shivdutt Singh, Alham~Fikri Aji, Jacki O{'}Neill, Ashutosh Modi, and Monojit Choudhury. 2024.
\newblock \href {https://doi.org/10.18653/v1/2024.emnlp-main.882} {Towards measuring and modeling {\textquotedblleft}culture{\textquotedblright} in {LLM}s: A survey}.
\newblock In \emph{Proceedings of the 2024 Conference on Empirical Methods in Natural Language Processing}, pages 15763--15784, Miami, Florida, USA. Association for Computational Linguistics.

\bibitem[{Agarwal et~al.(2024{\natexlab{a}})Agarwal, Naaman, and Vashistha}]{agarwal2024aisuggestionshomogenizewriting}
Dhruv Agarwal, Mor Naaman, and Aditya Vashistha. 2024{\natexlab{a}}.
\newblock \href {https://arxiv.org/abs/2409.11360} {Ai suggestions homogenize writing toward western styles and diminish cultural nuances}.
\newblock \emph{Preprint}, arXiv:2409.11360.

\bibitem[{Agarwal et~al.(2024{\natexlab{b}})Agarwal, Tanmay, Khandelwal, and Choudhury}]{agarwal2024ethicalreasoningmoralvalue}
Utkarsh Agarwal, Kumar Tanmay, Aditi Khandelwal, and Monojit Choudhury. 2024{\natexlab{b}}.
\newblock Ethical reasoning and moral value alignment of llms depend on the language we prompt them in.
\newblock In \emph{Proceedings of the 2024 Joint International Conference on Computational Linguistics, Language Resources and Evaluation (LREC-COLING 2024)}, pages 6330--6340.

\bibitem[{Aitchison(2005)}]{aitchison2005language}
Jean Aitchison. 2005.
\newblock Language change.
\newblock In \emph{The Routledge Companion to Semiotics and Linguistics}, pages 111--120. Routledge.

\bibitem[{Atkinson et~al.(2017)Atkinson, Baroni, Giacomin, Hunter, Prakken, Reed, Simari, Thimm, and Villata}]{atkinson2017towards}
Katie Atkinson, Pietro Baroni, Massimiliano Giacomin, Anthony Hunter, Henry Prakken, Chris Reed, Guillermo Simari, Matthias Thimm, and Serena Villata. 2017.
\newblock Towards artificial argumentation.
\newblock \emph{AI magazine}, 38(3):25--36.

\bibitem[{Bang et~al.(2007)Bang, Medin, and Atran}]{bang2007cultural}
Megan Bang, Douglas~L Medin, and Scott Atran. 2007.
\newblock Cultural mosaics and mental models of nature.
\newblock \emph{Proceedings of the National Academy of Sciences}, 104(35):13868--13874.

\bibitem[{Barnes(2013)}]{barnes2013three}
J.A. Barnes. 2013.
\newblock \href {https://books.google.ae/books?id=vSr-AQAAQBAJ} {\emph{Three Styles in the Study of Kinship}}.
\newblock Taylor \& Francis.

\bibitem[{Bella et~al.(2010)Bella, Ferri, Hern{\'a}ndez-Orallo, and Ram{\'\i}rez-Quintana}]{bella2010calibration}
Antonio Bella, C{\`e}sar Ferri, Jos{\'e} Hern{\'a}ndez-Orallo, and Mar{\'\i}a~Jos{\'e} Ram{\'\i}rez-Quintana. 2010.
\newblock Calibration of machine learning models.
\newblock In \emph{Handbook of Research on Machine Learning Applications and Trends: Algorithms, Methods, and Techniques}, pages 128--146. IGI Global.

\bibitem[{Bender and Beller(2013)}]{bender2013cognition}
Andrea Bender and Sieghard Beller. 2013.
\newblock Cognition is… fundamentally cultural.
\newblock \emph{Behavioral Sciences}, 3(1):42--54.

\bibitem[{Bender and Koller(2020)}]{bender-koller-2020-climbing}
Emily~M. Bender and Alexander Koller. 2020.
\newblock \href {https://doi.org/10.18653/v1/2020.acl-main.463} {Climbing towards {NLU}: {On} meaning, form, and understanding in the age of data}.
\newblock In \emph{Proceedings of the 58th Annual Meeting of the Association for Computational Linguistics}, pages 5185--5198, Online. Association for Computational Linguistics.

\bibitem[{Biesialska et~al.(2020)Biesialska, Biesialska, and Costa-Jussa}]{biesialska2020continual}
Magdalena Biesialska, Katarzyna Biesialska, and Marta~R Costa-Jussa. 2020.
\newblock Continual lifelong learning in natural language processing: A survey.
\newblock \emph{arXiv preprint arXiv:2012.09823}.

\bibitem[{Birukou et~al.(2013)Birukou, Blanzieri, Giorgini, and Giunchiglia}]{birukou2013formal}
Aliaksandr Birukou, Enrico Blanzieri, Paolo Giorgini, and Fausto Giunchiglia. 2013.
\newblock A formal definition of culture.
\newblock \emph{Models for intercultural collaboration and negotiation}, pages 1--26.

\bibitem[{Blake(2000)}]{blake2000defining}
Janet Blake. 2000.
\newblock On defining the cultural heritage.
\newblock \emph{International \& Comparative Law Quarterly}, 49(1):61--85.

\bibitem[{Bourdieu(1977)}]{Bourdieu_1977}
Pierre Bourdieu. 1977.
\newblock \emph{Outline of a Theory of Practice}.
\newblock Cambridge Studies in Social and Cultural Anthropology. Cambridge University Press.

\bibitem[{Boztemir and Çalışkan(2024)}]{Boztemir_2024}
Yiğithan Boztemir and Nilüfer Çalışkan. 2024.
\newblock \href {https://doi.org/10.36227/techrxiv.171504678.80895811/v1} {Analyzing and mitigating cultural hallucinations of commercial language models in turkish}.

\bibitem[{Bri{\~n}ol and DeMarree(2012)}]{brinol2012social}
Pablo Bri{\~n}ol and Kenneth~G DeMarree. 2012.
\newblock Social metacognition: Thinking about thinking in social psychology.
\newblock In \emph{Social metacognition}, pages 1--18. Psychology Press.

\bibitem[{Brinton(2005)}]{brinton2005lexicalization}
Laurel~J Brinton. 2005.
\newblock \emph{Lexicalization and language change}.
\newblock Cambridge University Press.

\bibitem[{Bybee(2015)}]{bybee2015language}
Joan Bybee. 2015.
\newblock \emph{Language change}.
\newblock Cambridge University Press.

\bibitem[{Cao et~al.(2023)Cao, Zhou, Lee, Cabello, Chen, and Hershcovich}]{cao-etal-2023-assessing}
Yong Cao, Li~Zhou, Seolhwa Lee, Laura Cabello, Min Chen, and Daniel Hershcovich. 2023.
\newblock \href {https://doi.org/10.18653/v1/2023.c3nlp-1.7} {Assessing cross-cultural alignment between {C}hat{GPT} and human societies: An empirical study}.
\newblock In \emph{Proceedings of the First Workshop on Cross-Cultural Considerations in NLP (C3NLP)}, pages 53--67, Dubrovnik, Croatia. Association for Computational Linguistics.

\bibitem[{Casper et~al.(2023)Casper, Davies, Shi, Gilbert, Scheurer, Rando, Freedman, Korbak, Lindner, Freire, Wang, Marks, Segerie, Carroll, Peng, Christoffersen, Damani, Slocum, Anwar, Siththaranjan, Nadeau, Michaud, Pfau, Krasheninnikov, Chen, Langosco, Hase, Biyik, Dragan, Krueger, Sadigh, and Hadfield-Menell}]{casper2023open}
Stephen Casper, Xander Davies, Claudia Shi, Thomas~Krendl Gilbert, J{\'e}r{\'e}my Scheurer, Javier Rando, Rachel Freedman, Tomek Korbak, David Lindner, Pedro Freire, Tony~Tong Wang, Samuel Marks, Charbel-Raphael Segerie, Micah Carroll, Andi Peng, Phillip~J.K. Christoffersen, Mehul Damani, Stewart Slocum, Usman Anwar, Anand Siththaranjan, Max Nadeau, Eric~J Michaud, Jacob Pfau, Dmitrii Krasheninnikov, Xin Chen, Lauro Langosco, Peter Hase, Erdem Biyik, Anca Dragan, David Krueger, Dorsa Sadigh, and Dylan Hadfield-Menell. 2023.
\newblock \href {https://openreview.net/forum?id=bx24KpJ4Eb} {Open problems and fundamental limitations of reinforcement learning from human feedback}.
\newblock \emph{Transactions on Machine Learning Research}.
\newblock Survey Certification, Featured Certification.

\bibitem[{Chaves and Gerosa(2021)}]{chaves2021should}
Ana~Paula Chaves and Marco~Aurelio Gerosa. 2021.
\newblock How should my chatbot interact? a survey on social characteristics in human--chatbot interaction design.
\newblock \emph{International Journal of Human--Computer Interaction}, 37(8):729--758.

\bibitem[{Chiu and Bendapudi(2012)}]{chiusocialmeta}
Chi~Yue Chiu and Namrita Bendapudi. 2012.
\newblock \href {https://doi.org/10.1037/a0030560} {Some thoughts on social metacognition.}
\newblock \emph{PsycCRITIQUES}, 57.

\bibitem[{Chiu et~al.(2024)Chiu, Jiang, Lin, Park, Li, Ravi, Bhatia, Antoniak, Tsvetkov, Shwartz, and Choi}]{chiu2024culturalbenchrobustdiversechallenging}
Yu~Ying Chiu, Liwei Jiang, Bill~Yuchen Lin, Chan~Young Park, Shuyue~Stella Li, Sahithya Ravi, Mehar Bhatia, Maria Antoniak, Yulia Tsvetkov, Vered Shwartz, and Yejin Choi. 2024.
\newblock \href {https://arxiv.org/abs/2410.02677} {Culturalbench: a robust, diverse and challenging benchmark on measuring the (lack of) cultural knowledge of llms}.
\newblock \emph{Preprint}, arXiv:2410.02677.

\bibitem[{Chomsky(2017)}]{CHOMSKY2017295}
Noam Chomsky. 2017.
\newblock \href {https://doi.org/10.1016/j.neubiorev.2017.01.053} {Language architecture and its import for evolution}.
\newblock \emph{Neuroscience \& Biobehavioral Reviews}, 81:295--300.
\newblock The Biology of Language.

\bibitem[{Chu et~al.(2024)Chu, Chen, Chen, Yu, He, Wang, Peng, Liu, Qin, and Liu}]{chu2023survey}
Zheng Chu, Jingchang Chen, Qianglong Chen, Weijiang Yu, Tao He, Haotian Wang, Weihua Peng, Ming Liu, Bing Qin, and Ting Liu. 2024.
\newblock \href {https://doi.org/10.18653/v1/2024.acl-long.65} {Navigate through enigmatic labyrinth a survey of chain of thought reasoning: Advances, frontiers and future}.
\newblock In \emph{Proceedings of the 62nd Annual Meeting of the Association for Computational Linguistics (Volume 1: Long Papers)}, pages 1173--1203, Bangkok, Thailand. Association for Computational Linguistics.

\bibitem[{Cialdini(2001)}]{cialdini2001science}
Robert~B Cialdini. 2001.
\newblock The science of persuasion.
\newblock \emph{Scientific American}, 284(2):76--81.

\bibitem[{Ciechanowski et~al.(2019)Ciechanowski, Przegalinska, Magnuski, and Gloor}]{ciechanowski2019shades}
Leon Ciechanowski, Aleksandra Przegalinska, Mikolaj Magnuski, and Peter Gloor. 2019.
\newblock In the shades of the uncanny valley: An experimental study of human--chatbot interaction.
\newblock \emph{Future Generation Computer Systems}, 92:539--548.

\bibitem[{Cohen(2009)}]{cohen2009many}
Adam~B Cohen. 2009.
\newblock Many forms of culture.
\newblock \emph{American psychologist}, 64(3):194.

\bibitem[{Cole and Packer(2019)}]{cole2019culture}
Michael Cole and Martin Packer. 2019.
\newblock Culture and cognition.
\newblock \emph{Cross-cultural psychology: Contemporary themes and perspectives}, pages 243--270.

\bibitem[{Collins and Gentner(1987)}]{collins1987people}
Allan Collins and Dedre Gentner. 1987.
\newblock How people construct mental models.
\newblock \emph{Cultural models in language and thought}, 243(1987):243--265.

\bibitem[{Conneau and Lample(2019)}]{conneau2019cross}
Alexis Conneau and Guillaume Lample. 2019.
\newblock Cross-lingual language model pretraining.
\newblock \emph{Advances in neural information processing systems}, 32.

\bibitem[{Cope(2022)}]{cope2022introduction}
Edward~Meredith Cope. 2022.
\newblock \emph{An introduction to Aristotle's rhetoric}.
\newblock BoD--Books on Demand.

\bibitem[{Dam et~al.(2024)Dam, Hong, Qiao, and Zhang}]{dam2024complete}
Sumit~Kumar Dam, Choong~Seon Hong, Yu~Qiao, and Chaoning Zhang. 2024.
\newblock \href {https://arxiv.org/abs/2406.16937} {A complete survey on llm-based ai chatbots}.
\newblock \emph{Preprint}, arXiv:2406.16937.

\bibitem[{Dammu et~al.(2024)Dammu, Jung, Singh, Choudhury, and Mitra}]{dammu2024theyunculturedunveilingcovert}
Preetam Prabhu~Srikar Dammu, Hayoung Jung, Anjali Singh, Monojit Choudhury, and Tanu Mitra. 2024.
\newblock \href {https://doi.org/10.18653/v1/2024.emnlp-main.1134} {{\textquotedblleft}they are uncultured{\textquotedblright}: Unveiling covert harms and social threats in {LLM} generated conversations}.
\newblock In \emph{Proceedings of the 2024 Conference on Empirical Methods in Natural Language Processing}, pages 20339--20369, Miami, Florida, USA. Association for Computational Linguistics.

\bibitem[{D'Andrade(1995)}]{d1995development}
R.G. D'Andrade. 1995.
\newblock \href {https://books.google.ae/books?id=2QCWe2r-pvwC} {\emph{The Development of Cognitive Anthropology}}.
\newblock The Development of Cognitive Anthropology. Cambridge University Press.

\bibitem[{Dawson et~al.(2024)Dawson, Mosunmola, Pocker, Dandekar, Dandekar, and Panat}]{dawson2024evaluatingculturalawarenessllms}
Fiifi Dawson, Zainab Mosunmola, Sahil Pocker, Raj~Abhijit Dandekar, Rajat Dandekar, and Sreedath Panat. 2024.
\newblock \href {https://arxiv.org/abs/2410.01811} {Evaluating cultural awareness of llms for yoruba, malayalam, and english}.
\newblock \emph{Preprint}, arXiv:2410.01811.

\bibitem[{Denzau et~al.(1994)Denzau, North et~al.}]{denzau1994shared}
Arthur~T Denzau, Douglass~C North, et~al. 1994.
\newblock Shared mental models: ideologies and institutions.
\newblock \emph{KYKLOS-BERNE-}, 47:3--3.

\bibitem[{Dong et~al.(2024)Dong, Li, Dai, Zheng, Ma, Li, Xia, Xu, Wu, Chang, Sun, Li, and Sui}]{dong2022survey}
Qingxiu Dong, Lei Li, Damai Dai, Ce~Zheng, Jingyuan Ma, Rui Li, Heming Xia, Jingjing Xu, Zhiyong Wu, Baobao Chang, Xu~Sun, Lei Li, and Zhifang Sui. 2024.
\newblock \href {https://doi.org/10.18653/v1/2024.emnlp-main.64} {A survey on in-context learning}.
\newblock In \emph{Proceedings of the 2024 Conference on Empirical Methods in Natural Language Processing}, pages 1107--1128, Miami, Florida, USA. Association for Computational Linguistics.

\bibitem[{Dubey et~al.(2024)Dubey, Jauhri, Pandey, Kadian, Al-Dahle, Letman, Mathur, Schelten, Yang, Fan, Goyal, Hartshorn, Yang, Mitra, Sravankumar, Korenev, Hinsvark, Rao, Zhang, Rodriguez, Gregerson, Spataru, Roziere, Biron, Tang, Chern, Caucheteux, Nayak, Bi, Marra, McConnell, Keller, Touret, Wu, Wong, Ferrer, Nikolaidis, Allonsius, Song, Pintz, Livshits, Esiobu, Choudhary, Mahajan, Garcia-Olano, Perino, Hupkes, Lakomkin, AlBadawy, Lobanova, Dinan, Smith, Radenovic, Zhang, Synnaeve, Lee, Anderson, Nail, Mialon, Pang, Cucurell, Nguyen, Korevaar, Xu, Touvron, Zarov, Ibarra, Kloumann, Misra, Evtimov, Copet, Lee, Geffert, Vranes, Park, Mahadeokar, Shah, van~der Linde, Billock, Hong, Lee, Fu, Chi, Huang, Liu, Wang, Yu, Bitton, Spisak, Park, Rocca, Johnstun, Saxe, Jia, Alwala, Upasani, Plawiak, Li, Heafield, Stone, El-Arini, Iyer, Malik, Chiu, Bhalla, Rantala-Yeary, van~der Maaten, Chen, Tan, Jenkins, Martin, Madaan, Malo, Blecher, Landzaat, de~Oliveira, Muzzi, Pasupuleti, Singh, Paluri, Kardas, Oldham, Rita,
  Pavlova, Kambadur, Lewis, Si, Singh, Hassan, Goyal, Torabi, Bashlykov, Bogoychev, Chatterji, Duchenne, Çelebi, Alrassy, Zhang, Li, Vasic, Weng, Bhargava, Dubal, Krishnan, Koura, Xu, He, Dong, Srinivasan, Ganapathy, Calderer, Cabral, Stojnic, Raileanu, Girdhar, Patel, Sauvestre, Polidoro, Sumbaly, Taylor, Silva, Hou, Wang, Hosseini, Chennabasappa, Singh, Bell, Kim, Edunov, Nie, Narang, Raparthy, Shen, Wan, Bhosale, Zhang, Vandenhende, Batra, Whitman, Sootla, Collot, Gururangan, Borodinsky, Herman, Fowler, Sheasha, Georgiou, Scialom, Speckbacher, Mihaylov, Xiao, Karn, Goswami, Gupta, Ramanathan, Kerkez, Gonguet, Do, Vogeti, Petrovic, Chu, Xiong, Fu, Meers, Martinet, Wang, Tan, Xie, Jia, Wang, Goldschlag, Gaur, Babaei, Wen, Song, Zhang, Li, Mao, Coudert, Yan, Chen, Papakipos, Singh, Grattafiori, Jain, Kelsey, Shajnfeld, Gangidi, Victoria, Goldstand, Menon, Sharma, Boesenberg, Vaughan, Baevski, Feinstein, Kallet, Sangani, Yunus, Lupu, Alvarado, Caples, Gu, Ho, Poulton, Ryan, Ramchandani, Franco, Saraf,
  Chowdhury, Gabriel, Bharambe, Eisenman, Yazdan, James, Maurer, Leonhardi, Huang, Loyd, Paola, Paranjape, Liu, Wu, Ni, Hancock, Wasti, Spence, Stojkovic, Gamido, Montalvo, Parker, Burton, Mejia, Wang, Kim, Zhou, Hu, Chu, Cai, Tindal, Feichtenhofer, Civin, Beaty, Kreymer, Li, Wyatt, Adkins, Xu, Testuggine, David, Parikh, Liskovich, Foss, Wang, Le, Holland, Dowling, Jamil, Montgomery, Presani, Hahn, Wood, Brinkman, Arcaute, Dunbar, Smothers, Sun, Kreuk, Tian, Ozgenel, Caggioni, Guzmán, Kanayet, Seide, Florez, Schwarz, Badeer, Swee, Halpern, Thattai, Herman, Sizov, Guangyi, Zhang, Lakshminarayanan, Shojanazeri, Zou, Wang, Zha, Habeeb, Rudolph, Suk, Aspegren, Goldman, Damlaj, Molybog, Tufanov, Veliche, Gat, Weissman, Geboski, Kohli, Asher, Gaya, Marcus, Tang, Chan, Zhen, Reizenstein, Teboul, Zhong, Jin, Yang, Cummings, Carvill, Shepard, McPhie, Torres, Ginsburg, Wang, Wu, U, Saxena, Prasad, Khandelwal, Zand, Matosich, Veeraraghavan, Michelena, Li, Huang, Chawla, Lakhotia, Huang, Chen, Garg, A, Silva, Bell,
  Zhang, Guo, Yu, Moshkovich, Wehrstedt, Khabsa, Avalani, Bhatt, Tsimpoukelli, Mankus, Hasson, Lennie, Reso, Groshev, Naumov, Lathi, Keneally, Seltzer, Valko, Restrepo, Patel, Vyatskov, Samvelyan, Clark, Macey, Wang, Hermoso, Metanat, Rastegari, Bansal, Santhanam, Parks, White, Bawa, Singhal, Egebo, Usunier, Laptev, Dong, Zhang, Cheng, Chernoguz, Hart, Salpekar, Kalinli, Kent, Parekh, Saab, Balaji, Rittner, Bontrager, Roux, Dollar, Zvyagina, Ratanchandani, Yuvraj, Liang, Alao, Rodriguez, Ayub, Murthy, Nayani, Mitra, Li, Hogan, Battey, Wang, Maheswari, Howes, Rinott, Bondu, Datta, Chugh, Hunt, Dhillon, Sidorov, Pan, Verma, Yamamoto, Ramaswamy, Lindsay, Lindsay, Feng, Lin, Zha, Shankar, Zhang, Zhang, Wang, Agarwal, Sajuyigbe, Chintala, Max, Chen, Kehoe, Satterfield, Govindaprasad, Gupta, Cho, Virk, Subramanian, Choudhury, Goldman, Remez, Glaser, Best, Kohler, Robinson, Li, Zhang, Matthews, Chou, Shaked, Vontimitta, Ajayi, Montanez, Mohan, Kumar, Mangla, Albiero, Ionescu, Poenaru, Mihailescu, Ivanov, Li, Wang,
  Jiang, Bouaziz, Constable, Tang, Wang, Wu, Wang, Xia, Wu, Gao, Chen, Hu, Jia, Qi, Li, Zhang, Zhang, Adi, Nam, Yu, Wang, Hao, Qian, He, Rait, DeVito, Rosnbrick, Wen, Yang, and Zhao}]{dubey2024llama}
Abhimanyu Dubey, Abhinav Jauhri, Abhinav Pandey, Abhishek Kadian, Ahmad Al-Dahle, Aiesha Letman, Akhil Mathur, Alan Schelten, Amy Yang, Angela Fan, Anirudh Goyal, Anthony Hartshorn, Aobo Yang, Archi Mitra, Archie Sravankumar, Artem Korenev, Arthur Hinsvark, Arun Rao, Aston Zhang, Aurelien Rodriguez, Austen Gregerson, Ava Spataru, Baptiste Roziere, Bethany Biron, Binh Tang, Bobbie Chern, Charlotte Caucheteux, Chaya Nayak, Chloe Bi, Chris Marra, Chris McConnell, Christian Keller, Christophe Touret, Chunyang Wu, Corinne Wong, Cristian~Canton Ferrer, Cyrus Nikolaidis, Damien Allonsius, Daniel Song, Danielle Pintz, Danny Livshits, David Esiobu, Dhruv Choudhary, Dhruv Mahajan, Diego Garcia-Olano, Diego Perino, Dieuwke Hupkes, Egor Lakomkin, Ehab AlBadawy, Elina Lobanova, Emily Dinan, Eric~Michael Smith, Filip Radenovic, Frank Zhang, Gabriel Synnaeve, Gabrielle Lee, Georgia~Lewis Anderson, Graeme Nail, Gregoire Mialon, Guan Pang, Guillem Cucurell, Hailey Nguyen, Hannah Korevaar, Hu~Xu, Hugo Touvron, Iliyan Zarov,
  Imanol~Arrieta Ibarra, Isabel Kloumann, Ishan Misra, Ivan Evtimov, Jade Copet, Jaewon Lee, Jan Geffert, Jana Vranes, Jason Park, Jay Mahadeokar, Jeet Shah, Jelmer van~der Linde, Jennifer Billock, Jenny Hong, Jenya Lee, Jeremy Fu, Jianfeng Chi, Jianyu Huang, Jiawen Liu, Jie Wang, Jiecao Yu, Joanna Bitton, Joe Spisak, Jongsoo Park, Joseph Rocca, Joshua Johnstun, Joshua Saxe, Junteng Jia, Kalyan~Vasuden Alwala, Kartikeya Upasani, Kate Plawiak, Ke~Li, Kenneth Heafield, Kevin Stone, Khalid El-Arini, Krithika Iyer, Kshitiz Malik, Kuenley Chiu, Kunal Bhalla, Lauren Rantala-Yeary, Laurens van~der Maaten, Lawrence Chen, Liang Tan, Liz Jenkins, Louis Martin, Lovish Madaan, Lubo Malo, Lukas Blecher, Lukas Landzaat, Luke de~Oliveira, Madeline Muzzi, Mahesh Pasupuleti, Mannat Singh, Manohar Paluri, Marcin Kardas, Mathew Oldham, Mathieu Rita, Maya Pavlova, Melanie Kambadur, Mike Lewis, Min Si, Mitesh~Kumar Singh, Mona Hassan, Naman Goyal, Narjes Torabi, Nikolay Bashlykov, Nikolay Bogoychev, Niladri Chatterji, Olivier
  Duchenne, Onur Çelebi, Patrick Alrassy, Pengchuan Zhang, Pengwei Li, Petar Vasic, Peter Weng, Prajjwal Bhargava, Pratik Dubal, Praveen Krishnan, Punit~Singh Koura, Puxin Xu, Qing He, Qingxiao Dong, Ragavan Srinivasan, Raj Ganapathy, Ramon Calderer, Ricardo~Silveira Cabral, Robert Stojnic, Roberta Raileanu, Rohit Girdhar, Rohit Patel, Romain Sauvestre, Ronnie Polidoro, Roshan Sumbaly, Ross Taylor, Ruan Silva, Rui Hou, Rui Wang, Saghar Hosseini, Sahana Chennabasappa, Sanjay Singh, Sean Bell, Seohyun~Sonia Kim, Sergey Edunov, Shaoliang Nie, Sharan Narang, Sharath Raparthy, Sheng Shen, Shengye Wan, Shruti Bhosale, Shun Zhang, Simon Vandenhende, Soumya Batra, Spencer Whitman, Sten Sootla, Stephane Collot, Suchin Gururangan, Sydney Borodinsky, Tamar Herman, Tara Fowler, Tarek Sheasha, Thomas Georgiou, Thomas Scialom, Tobias Speckbacher, Todor Mihaylov, Tong Xiao, Ujjwal Karn, Vedanuj Goswami, Vibhor Gupta, Vignesh Ramanathan, Viktor Kerkez, Vincent Gonguet, Virginie Do, Vish Vogeti, Vladan Petrovic, Weiwei Chu,
  Wenhan Xiong, Wenyin Fu, Whitney Meers, Xavier Martinet, Xiaodong Wang, Xiaoqing~Ellen Tan, Xinfeng Xie, Xuchao Jia, Xuewei Wang, Yaelle Goldschlag, Yashesh Gaur, Yasmine Babaei, Yi~Wen, Yiwen Song, Yuchen Zhang, Yue Li, Yuning Mao, Zacharie~Delpierre Coudert, Zheng Yan, Zhengxing Chen, Zoe Papakipos, Aaditya Singh, Aaron Grattafiori, Abha Jain, Adam Kelsey, Adam Shajnfeld, Adithya Gangidi, Adolfo Victoria, Ahuva Goldstand, Ajay Menon, Ajay Sharma, Alex Boesenberg, Alex Vaughan, Alexei Baevski, Allie Feinstein, Amanda Kallet, Amit Sangani, Anam Yunus, Andrei Lupu, Andres Alvarado, Andrew Caples, Andrew Gu, Andrew Ho, Andrew Poulton, Andrew Ryan, Ankit Ramchandani, Annie Franco, Aparajita Saraf, Arkabandhu Chowdhury, Ashley Gabriel, Ashwin Bharambe, Assaf Eisenman, Azadeh Yazdan, Beau James, Ben Maurer, Benjamin Leonhardi, Bernie Huang, Beth Loyd, Beto~De Paola, Bhargavi Paranjape, Bing Liu, Bo~Wu, Boyu Ni, Braden Hancock, Bram Wasti, Brandon Spence, Brani Stojkovic, Brian Gamido, Britt Montalvo, Carl
  Parker, Carly Burton, Catalina Mejia, Changhan Wang, Changkyu Kim, Chao Zhou, Chester Hu, Ching-Hsiang Chu, Chris Cai, Chris Tindal, Christoph Feichtenhofer, Damon Civin, Dana Beaty, Daniel Kreymer, Daniel Li, Danny Wyatt, David Adkins, David Xu, Davide Testuggine, Delia David, Devi Parikh, Diana Liskovich, Didem Foss, Dingkang Wang, Duc Le, Dustin Holland, Edward Dowling, Eissa Jamil, Elaine Montgomery, Eleonora Presani, Emily Hahn, Emily Wood, Erik Brinkman, Esteban Arcaute, Evan Dunbar, Evan Smothers, Fei Sun, Felix Kreuk, Feng Tian, Firat Ozgenel, Francesco Caggioni, Francisco Guzmán, Frank Kanayet, Frank Seide, Gabriela~Medina Florez, Gabriella Schwarz, Gada Badeer, Georgia Swee, Gil Halpern, Govind Thattai, Grant Herman, Grigory Sizov, Guangyi, Zhang, Guna Lakshminarayanan, Hamid Shojanazeri, Han Zou, Hannah Wang, Hanwen Zha, Haroun Habeeb, Harrison Rudolph, Helen Suk, Henry Aspegren, Hunter Goldman, Ibrahim Damlaj, Igor Molybog, Igor Tufanov, Irina-Elena Veliche, Itai Gat, Jake Weissman, James
  Geboski, James Kohli, Japhet Asher, Jean-Baptiste Gaya, Jeff Marcus, Jeff Tang, Jennifer Chan, Jenny Zhen, Jeremy Reizenstein, Jeremy Teboul, Jessica Zhong, Jian Jin, Jingyi Yang, Joe Cummings, Jon Carvill, Jon Shepard, Jonathan McPhie, Jonathan Torres, Josh Ginsburg, Junjie Wang, Kai Wu, Kam~Hou U, Karan Saxena, Karthik Prasad, Kartikay Khandelwal, Katayoun Zand, Kathy Matosich, Kaushik Veeraraghavan, Kelly Michelena, Keqian Li, Kun Huang, Kunal Chawla, Kushal Lakhotia, Kyle Huang, Lailin Chen, Lakshya Garg, Lavender A, Leandro Silva, Lee Bell, Lei Zhang, Liangpeng Guo, Licheng Yu, Liron Moshkovich, Luca Wehrstedt, Madian Khabsa, Manav Avalani, Manish Bhatt, Maria Tsimpoukelli, Martynas Mankus, Matan Hasson, Matthew Lennie, Matthias Reso, Maxim Groshev, Maxim Naumov, Maya Lathi, Meghan Keneally, Michael~L. Seltzer, Michal Valko, Michelle Restrepo, Mihir Patel, Mik Vyatskov, Mikayel Samvelyan, Mike Clark, Mike Macey, Mike Wang, Miquel~Jubert Hermoso, Mo~Metanat, Mohammad Rastegari, Munish Bansal, Nandhini
  Santhanam, Natascha Parks, Natasha White, Navyata Bawa, Nayan Singhal, Nick Egebo, Nicolas Usunier, Nikolay~Pavlovich Laptev, Ning Dong, Ning Zhang, Norman Cheng, Oleg Chernoguz, Olivia Hart, Omkar Salpekar, Ozlem Kalinli, Parkin Kent, Parth Parekh, Paul Saab, Pavan Balaji, Pedro Rittner, Philip Bontrager, Pierre Roux, Piotr Dollar, Polina Zvyagina, Prashant Ratanchandani, Pritish Yuvraj, Qian Liang, Rachad Alao, Rachel Rodriguez, Rafi Ayub, Raghotham Murthy, Raghu Nayani, Rahul Mitra, Raymond Li, Rebekkah Hogan, Robin Battey, Rocky Wang, Rohan Maheswari, Russ Howes, Ruty Rinott, Sai~Jayesh Bondu, Samyak Datta, Sara Chugh, Sara Hunt, Sargun Dhillon, Sasha Sidorov, Satadru Pan, Saurabh Verma, Seiji Yamamoto, Sharadh Ramaswamy, Shaun Lindsay, Shaun Lindsay, Sheng Feng, Shenghao Lin, Shengxin~Cindy Zha, Shiva Shankar, Shuqiang Zhang, Shuqiang Zhang, Sinong Wang, Sneha Agarwal, Soji Sajuyigbe, Soumith Chintala, Stephanie Max, Stephen Chen, Steve Kehoe, Steve Satterfield, Sudarshan Govindaprasad, Sumit Gupta,
  Sungmin Cho, Sunny Virk, Suraj Subramanian, Sy~Choudhury, Sydney Goldman, Tal Remez, Tamar Glaser, Tamara Best, Thilo Kohler, Thomas Robinson, Tianhe Li, Tianjun Zhang, Tim Matthews, Timothy Chou, Tzook Shaked, Varun Vontimitta, Victoria Ajayi, Victoria Montanez, Vijai Mohan, Vinay~Satish Kumar, Vishal Mangla, Vítor Albiero, Vlad Ionescu, Vlad Poenaru, Vlad~Tiberiu Mihailescu, Vladimir Ivanov, Wei Li, Wenchen Wang, Wenwen Jiang, Wes Bouaziz, Will Constable, Xiaocheng Tang, Xiaofang Wang, Xiaojian Wu, Xiaolan Wang, Xide Xia, Xilun Wu, Xinbo Gao, Yanjun Chen, Ye~Hu, Ye~Jia, Ye~Qi, Yenda Li, Yilin Zhang, Ying Zhang, Yossi Adi, Youngjin Nam, Yu, Wang, Yuchen Hao, Yundi Qian, Yuzi He, Zach Rait, Zachary DeVito, Zef Rosnbrick, Zhaoduo Wen, Zhenyu Yang, and Zhiwei Zhao. 2024.
\newblock \href {https://arxiv.org/abs/2407.21783} {The llama 3 herd of models}.
\newblock \emph{Preprint}, arXiv:2407.21783.

\bibitem[{Dudy et~al.(2024)Dudy, Ahmad, Kitajima, and Lapedriza}]{dudy2024analyzing}
Shiran Dudy, Ibrahim~Said Ahmad, Ryoko Kitajima, and Agata Lapedriza. 2024.
\newblock Analyzing cultural representations of emotions in llms through mixed emotion survey.
\newblock \emph{arXiv preprint arXiv:2408.02143}.

\bibitem[{Eckert(2012)}]{eckert2012three}
Penelope Eckert. 2012.
\newblock Three waves of variation study: The emergence of meaning in the study of sociolinguistic variation.
\newblock \emph{Annual review of Anthropology}, 41(1):87--100.

\bibitem[{Eckert and Rickford(2001)}]{eckert2001style}
Penelope Eckert and John~R Rickford. 2001.
\newblock \emph{Style and sociolinguistic variation}.
\newblock Cambridge University Press.

\bibitem[{Fan et~al.(2024)Fan, Ding, Ning, Wang, Li, Yin, Chua, and Li}]{fan2024survey}
Wenqi Fan, Yujuan Ding, Liangbo Ning, Shijie Wang, Hengyun Li, Dawei Yin, Tat-Seng Chua, and Qing Li. 2024.
\newblock A survey on rag meeting llms: Towards retrieval-augmented large language models.
\newblock In \emph{Proceedings of the 30th ACM SIGKDD Conference on Knowledge Discovery and Data Mining}, pages 6491--6501.

\bibitem[{Ferraro(2019)}]{10.1145/3298689.3347052}
Andres Ferraro. 2019.
\newblock \href {https://doi.org/10.1145/3298689.3347052} {Music cold-start and long-tail recommendation: bias in deep representations}.
\newblock In \emph{Proceedings of the 13th ACM Conference on Recommender Systems}, RecSys '19, page 586–590, New York, NY, USA. Association for Computing Machinery.

\bibitem[{Fitch et~al.(2005)Fitch, Hauser, and Chomsky}]{FITCH2005179}
W.~Tecumseh Fitch, Marc~D. Hauser, and Noam Chomsky. 2005.
\newblock \href {https://doi.org/10.1016/j.cognition.2005.02.005} {The evolution of the language faculty: Clarifications and implications}.
\newblock \emph{Cognition}, 97(2):179--210.

\bibitem[{Fr{\"o}hlich et~al.(2019)Fr{\"o}hlich, Sievers, Townsend, Gruber, and van Schaik}]{frohlich2019multimodal}
Marlen Fr{\"o}hlich, Christine Sievers, Simon~W Townsend, Thibaud Gruber, and Carel~P van Schaik. 2019.
\newblock Multimodal communication and language origins: integrating gestures and vocalizations.
\newblock \emph{Biological Reviews}, 94(5):1809--1829.

\bibitem[{Fu et~al.(2024)Fu, Bailis, Stoica, and Zhang}]{fu2024break}
Yichao Fu, Peter Bailis, Ion Stoica, and Hao Zhang. 2024.
\newblock Break the sequential dependency of llm inference using lookahead decoding.
\newblock In \emph{Proceedings of the 41st International Conference on Machine Learning}, ICML'24. JMLR.org.

\bibitem[{Gao et~al.(2023)Gao, Xiong, Gao, Jia, Pan, Bi, Dai, Sun, and Wang}]{gao2023retrieval}
Yunfan Gao, Yun Xiong, Xinyu Gao, Kangxiang Jia, Jinliu Pan, Yuxi Bi, Yi~Dai, Jiawei Sun, and Haofen Wang. 2023.
\newblock Retrieval-augmented generation for large language models: A survey.
\newblock \emph{arXiv preprint arXiv:2312.10997}.

\bibitem[{Geert and Hofstede(2004)}]{book2}
Hofstede Geert and Gert~Jan Hofstede. 2004.
\newblock \emph{Cultures and Organizations. Software of the Mind}, volume~2.

\bibitem[{Geertz(1973)}]{geertz2017interpretation}
Clifford Geertz. 1973.
\newblock \emph{The interpretation of cultures}.
\newblock Basic books.

\bibitem[{Grieve et~al.(2025)Grieve, Bartl, Fuoli, Grafmiller, Huang, Jawerbaum, Murakami, Perlman, Roemling, and Winter}]{grieve2024sociolinguisticfoundationslanguagemodeling}
Jack Grieve, Sara Bartl, Matteo Fuoli, Jason Grafmiller, Weihang Huang, Alejandro Jawerbaum, Akira Murakami, Marcus Perlman, Dana Roemling, and Bodo Winter. 2025.
\newblock The sociolinguistic foundations of language modeling.
\newblock \emph{Frontiers in Artificial Intelligence}, 7:1472411.

\bibitem[{Griffith et~al.(2013)Griffith, Subramanian, Scholz, Isbell, and Thomaz}]{griffith2013policy}
Shane Griffith, Kaushik Subramanian, Jonathan Scholz, Charles~L Isbell, and Andrea~L Thomaz. 2013.
\newblock Policy shaping: Integrating human feedback with reinforcement learning.
\newblock \emph{Advances in neural information processing systems}, 26.

\bibitem[{Haffari et~al.(2017)Haffari, Tran, and Carman}]{haffari-etal-2017-efficient}
Gholamreza Haffari, Tuan~Dung Tran, and Mark Carman. 2017.
\newblock \href {https://aclanthology.org/E17-1039} {Efficient benchmarking of {NLP} {API}s using multi-armed bandits}.
\newblock In \emph{Proceedings of the 15th Conference of the {E}uropean Chapter of the Association for Computational Linguistics: Volume 1, Long Papers}, pages 408--416, Valencia, Spain. Association for Computational Linguistics.

\bibitem[{Harris(1951)}]{harris1951methods}
Zellig Harris. 1951.
\newblock Methods in structural linguistics.

\bibitem[{Harris(1963)}]{harris1963structural}
Zellig~Sabbettai Harris. 1963.
\newblock Structural linguistics.

\bibitem[{Hauser et~al.(2014)Hauser, Yang, Berwick, Tattersall, Ryan, Watumull, Chomsky, and Lewontin}]{10.3389/fpsyg.2014.00401}
Marc~D. Hauser, Charles Yang, Robert~C. Berwick, Ian Tattersall, Michael~J. Ryan, Jeffrey Watumull, Noam Chomsky, and Richard~C. Lewontin. 2014.
\newblock \href {https://doi.org/10.3389/fpsyg.2014.00401} {The mystery of language evolution}.
\newblock \emph{Frontiers in Psychology}, 5.

\bibitem[{H{\'e}naff(1998)}]{henaff1998claude}
Marcel H{\'e}naff. 1998.
\newblock \emph{Claude Levi-Strauss and the making of structural anthropology}.
\newblock U of Minnesota Press.

\bibitem[{Henrich et~al.(2010)Henrich, Heine, and Norenzayan}]{henrich2010weirdest}
Joseph Henrich, Steven~J Heine, and Ara Norenzayan. 2010.
\newblock The weirdest people in the world?
\newblock \emph{Behavioral and brain sciences}, 33(2-3):61--83.

\bibitem[{Hofstede(2001)}]{book1}
Geert Hofstede. 2001.
\newblock \href {https://doi.org/10.1016/S0005-7967(02)00184-5} {\emph{Culture's Consequences: Comparing Values, Behaviors, Institutions and Organizations Across Nations}}, volume~41.

\bibitem[{Hospedales et~al.(2021)Hospedales, Antoniou, Micaelli, and Storkey}]{hospedales2021meta}
Timothy Hospedales, Antreas Antoniou, Paul Micaelli, and Amos Storkey. 2021.
\newblock Meta-learning in neural networks: A survey.
\newblock \emph{IEEE transactions on pattern analysis and machine intelligence}, 44(9):5149--5169.

\bibitem[{Hu et~al.(2008)Hu, Koren, and Volinsky}]{hu2008collaborative}
Yifan Hu, Yehuda Koren, and Chris Volinsky. 2008.
\newblock Collaborative filtering for implicit feedback datasets.
\newblock In \emph{2008 Eighth IEEE international conference on data mining}, pages 263--272. Ieee.

\bibitem[{Huang et~al.(2025)Huang, Yu, Ma, Zhong, Feng, Wang, Chen, Peng, Feng, Qin, and Liu}]{huang2023survey}
Lei Huang, Weijiang Yu, Weitao Ma, Weihong Zhong, Zhangyin Feng, Haotian Wang, Qianglong Chen, Weihua Peng, Xiaocheng Feng, Bing Qin, and Ting Liu. 2025.
\newblock \href {https://doi.org/10.1145/3703155} {A survey on hallucination in large language models: Principles, taxonomy, challenges, and open questions}.
\newblock \emph{ACM Trans. Inf. Syst.}, 43(2).

\bibitem[{Jahoda and Lewis(2015)}]{jahoda2015acquiring}
Gustav Jahoda and Ioan Lewis. 2015.
\newblock \emph{Acquiring Culture (Psychology Revivals): Cross Cultural Studies in Child Development}.
\newblock Psychology Press.

\bibitem[{Jha et~al.(2023)Jha, Mostafazadeh~Davani, Reddy, Dave, Prabhakaran, and Dev}]{jha-etal-2023-seegull}
Akshita Jha, Aida Mostafazadeh~Davani, Chandan~K Reddy, Shachi Dave, Vinodkumar Prabhakaran, and Sunipa Dev. 2023.
\newblock \href {https://doi.org/10.18653/v1/2023.acl-long.548} {{S}ee{GULL}: A stereotype benchmark with broad geo-cultural coverage leveraging generative models}.
\newblock In \emph{Proceedings of the 61st Annual Meeting of the Association for Computational Linguistics (Volume 1: Long Papers)}, pages 9851--9870, Toronto, Canada. Association for Computational Linguistics.

\bibitem[{Ji et~al.(2023)Ji, Lee, Frieske, Yu, Su, Xu, Ishii, Bang, Madotto, and Fung}]{hallucination}
Ziwei Ji, Nayeon Lee, Rita Frieske, Tiezheng Yu, Dan Su, Yan Xu, Etsuko Ishii, Ye~Jin Bang, Andrea Madotto, and Pascale Fung. 2023.
\newblock \href {https://doi.org/10.1145/3571730} {Survey of hallucination in natural language generation}.
\newblock \emph{ACM Comput. Surv.}, 55(12).

\bibitem[{Jin et~al.(2024)Jin, Kim, Lee, Yoo, Oh, and Lee}]{jin2024kobbqkoreanbiasbenchmark}
Jiho Jin, Jiseon Kim, Nayeon Lee, Haneul Yoo, Alice Oh, and Hwaran Lee. 2024.
\newblock \href {https://doi.org/10.1162/tacl_a_00661} {{K}o{BBQ}: {K}orean bias benchmark for question answering}.
\newblock \emph{Transactions of the Association for Computational Linguistics}, 12:507--524.

\bibitem[{Jonassen and Henning(1999)}]{jonassen1999mental}
David~H Jonassen and Philip Henning. 1999.
\newblock Mental models: Knowledge in the head and knowledge in the world.
\newblock \emph{Educational technology}, pages 37--42.

\bibitem[{Keller(1994)}]{keller1994language}
Rudi Keller. 1994.
\newblock \emph{On language change: The invisible hand in language}.
\newblock Psychology Press.

\bibitem[{Kharchenko et~al.(2024{\natexlab{a}})Kharchenko, Roosta, Chadha, and Shah}]{kharchenko2024well}
Julia Kharchenko, Tanya Roosta, Aman Chadha, and Chirag Shah. 2024{\natexlab{a}}.
\newblock How well do llms represent values across cultures? empirical analysis of llm responses based on hofstede cultural dimensions.
\newblock \emph{arXiv preprint arXiv:2406.14805}.

\bibitem[{Kharchenko et~al.(2024{\natexlab{b}})Kharchenko, Roosta, Chadha, and Shah}]{kharchenko2024llmsrepresentvaluescultures}
Julia Kharchenko, Tanya Roosta, Aman Chadha, and Chirag Shah. 2024{\natexlab{b}}.
\newblock \href {https://arxiv.org/abs/2406.14805} {How well do llms represent values across cultures? empirical analysis of llm responses based on hofstede cultural dimensions}.
\newblock \emph{Preprint}, arXiv:2406.14805.

\bibitem[{Kim et~al.(2017)Kim, Kim, Sarikaya, and Fosler-Lussier}]{kim2017cross}
Joo-Kyung Kim, Young-Bum Kim, Ruhi Sarikaya, and Eric Fosler-Lussier. 2017.
\newblock Cross-lingual transfer learning for pos tagging without cross-lingual resources.
\newblock In \emph{Proceedings of the 2017 conference on empirical methods in natural language processing}, pages 2832--2838.

\bibitem[{Koto et~al.(2023)Koto, Aisyah, Li, and Baldwin}]{koto2023largelanguagemodelspass}
Fajri Koto, Nurul Aisyah, Haonan Li, and Timothy Baldwin. 2023.
\newblock \href {https://doi.org/10.18653/v1/2023.emnlp-main.760} {Large language models only pass primary school exams in {I}ndonesia: A comprehensive test on {I}ndo{MMLU}}.
\newblock In \emph{Proceedings of the 2023 Conference on Empirical Methods in Natural Language Processing}, pages 12359--12374, Singapore. Association for Computational Linguistics.

\bibitem[{Koto et~al.(2024)Koto, Mahendra, Aisyah, and Baldwin}]{koto2024indocultureexploringgeographicallyinfluencedcultural}
Fajri Koto, Rahmad Mahendra, Nurul Aisyah, and Timothy Baldwin. 2024.
\newblock \href {https://doi.org/10.1162/tacl_a_00726} {{I}ndo{C}ulture: Exploring geographically influenced cultural commonsense reasoning across eleven {I}ndonesian provinces}.
\newblock \emph{Transactions of the Association for Computational Linguistics}, 12:1703--1719.

\bibitem[{Kovač et~al.(2023)Kovač, Sawayama, Portelas, Colas, Dominey, and Oudeyer}]{kovač2023largelanguagemodelssuperpositions}
Grgur Kovač, Masataka Sawayama, Rémy Portelas, Cédric Colas, Peter~Ford Dominey, and Pierre-Yves Oudeyer. 2023.
\newblock \href {https://arxiv.org/abs/2307.07870} {Large language models as superpositions of cultural perspectives}.
\newblock \emph{Preprint}, arXiv:2307.07870.

\bibitem[{Kuper(1988)}]{kuper1988invention}
A.~Kuper. 1988.
\newblock \href {https://books.google.ae/books?id=yg4okTJNE24C} {\emph{The Invention of Primitive Society: Transformations of an Illusion}}.
\newblock Routledge.

\bibitem[{LeCompte and Schensul(2010)}]{lecompte2010designing}
Margaret~Diane LeCompte and Jean~J Schensul. 2010.
\newblock \emph{Designing \& conducting ethnographic research: An introduction}, volume~1.
\newblock Rowman Altamira.

\bibitem[{Lent et~al.(2024)Lent, Tatariya, Dabre, Chen, Fekete, Ploeger, Zhou, Armstrong, Eijansantos, Malau, Heje, Lavrinovics, Kanojia, Belony, Bollmann, Grobol, Lhoneux, Hershcovich, DeGraff, Søgaard, and Bjerva}]{10.1162/tacl_a_00682}
Heather Lent, Kushal Tatariya, Raj Dabre, Yiyi Chen, Marcell Fekete, Esther Ploeger, Li~Zhou, Ruth-Ann Armstrong, Abee Eijansantos, Catriona Malau, Hans~Erik Heje, Ernests Lavrinovics, Diptesh Kanojia, Paul Belony, Marcel Bollmann, Loïc Grobol, Miryam~de Lhoneux, Daniel Hershcovich, Michel DeGraff, Anders Søgaard, and Johannes Bjerva. 2024.
\newblock \href {https://doi.org/10.1162/tacl_a_00682} {{CreoleVal: Multilingual Multitask Benchmarks for Creoles}}.
\newblock \emph{Transactions of the Association for Computational Linguistics}, 12:950--978.

\bibitem[{Leung et~al.(2013)Leung, Lee, and Chiu}]{leung2013meta}
Angela K-y Leung, Sau-lai Lee, and Chi-yue Chiu. 2013.
\newblock Meta-knowledge of culture promotes cultural competence.
\newblock \emph{Journal of Cross-Cultural Psychology}, 44(6):992--1006.

\bibitem[{Lewis et~al.(2020)Lewis, Perez, Piktus, Petroni, Karpukhin, Goyal, K{\"u}ttler, Lewis, Yih, Rockt{\"a}schel et~al.}]{lewis2020retrieval}
Patrick Lewis, Ethan Perez, Aleksandra Piktus, Fabio Petroni, Vladimir Karpukhin, Naman Goyal, Heinrich K{\"u}ttler, Mike Lewis, Wen-tau Yih, Tim Rockt{\"a}schel, et~al. 2020.
\newblock Retrieval-augmented generation for knowledge-intensive nlp tasks.
\newblock \emph{Advances in Neural Information Processing Systems}, 33:9459--9474.

\bibitem[{LI et~al.(2024{\natexlab{a}})LI, Chen, Wang, Sitaram, and Xie}]{li2024culturellmincorporatingculturaldifferences}
CHENG LI, Mengzhuo Chen, Jindong Wang, Sunayana Sitaram, and Xing Xie. 2024{\natexlab{a}}.
\newblock \href {https://openreview.net/forum?id=sIsbOkQmBL} {Culture{LLM}: Incorporating cultural differences into large language models}.
\newblock In \emph{The Thirty-eighth Annual Conference on Neural Information Processing Systems}.

\bibitem[{LI et~al.(2024{\natexlab{b}})LI, Teney, Yang, Wen, Xie, and Wang}]{li2024culturepark}
CHENG LI, Damien Teney, Linyi Yang, Qingsong Wen, Xing Xie, and Jindong Wang. 2024{\natexlab{b}}.
\newblock \href {https://openreview.net/forum?id=bIFHHf2RoD} {Culturepark: Boosting cross-cultural understanding in large language models}.
\newblock In \emph{The Thirty-eighth Annual Conference on Neural Information Processing Systems}.

\bibitem[{Li et~al.(2022)Li, Su, Cai, Wang, and Liu}]{li2022survey}
Huayang Li, Yixuan Su, Deng Cai, Yan Wang, and Lemao Liu. 2022.
\newblock A survey on retrieval-augmented text generation.
\newblock \emph{arXiv preprint arXiv:2202.01110}.

\bibitem[{Li et~al.(2024)Li, Jiang, Dziri, Ren, and Choi}]{li2024culturegenrevealingglobalcultural}
Huihan Li, Liwei Jiang, Nouha Dziri, Xiang Ren, and Yejin Choi. 2024.
\newblock \href {https://openreview.net/forum?id=DbsLm2KAqP} {{CULTURE}-{GEN}: Revealing global cultural perception in language models through natural language prompting}.
\newblock In \emph{First Conference on Language Modeling}.

\bibitem[{Lichtenberg et~al.(2024)Lichtenberg, Buchholz, and Schwöbel}]{lichtenberg2024largelanguagemodelsrecommender}
Jan~Malte Lichtenberg, Alexander Buchholz, and Pola Schwöbel. 2024.
\newblock \href {https://arxiv.org/abs/2406.01285} {Large language models as recommender systems: A study of popularity bias}.
\newblock \emph{Preprint}, arXiv:2406.01285.

\bibitem[{Lightfoot(2002)}]{lightfoot2002explaining}
David~W Lightfoot. 2002.
\newblock Explaining language change: An evolutionary approach.

\bibitem[{Liu and Mazumder(2020)}]{liu2020lifelong}
Bing Liu and Sahisnu Mazumder. 2020.
\newblock Lifelong learning dialogue systems: Chatbots that self-learn on the job.
\newblock \emph{arXiv preprint arXiv:2009.10750}.

\bibitem[{Liu et~al.(2024)Liu, Koto, Baldwin, and Gurevych}]{liu2024multilingualllmsculturallydiversereasoners}
Chen Liu, Fajri Koto, Timothy Baldwin, and Iryna Gurevych. 2024.
\newblock \href {https://doi.org/10.18653/v1/2024.naacl-long.112} {Are multilingual {LLM}s culturally-diverse reasoners? an investigation into multicultural proverbs and sayings}.
\newblock In \emph{Proceedings of the 2024 Conference of the North American Chapter of the Association for Computational Linguistics: Human Language Technologies (Volume 1: Long Papers)}, pages 2016--2039, Mexico City, Mexico. Association for Computational Linguistics.

\bibitem[{Lu et~al.(2010)Lu, Pal, and Pal}]{pmlr-v9-lu10a}
Tyler Lu, David Pal, and Martin Pal. 2010.
\newblock \href {https://proceedings.mlr.press/v9/lu10a.html} {Contextual multi-armed bandits}.
\newblock In \emph{Proceedings of the Thirteenth International Conference on Artificial Intelligence and Statistics}, volume~9 of \emph{Proceedings of Machine Learning Research}, pages 485--492, Chia Laguna Resort, Sardinia, Italy. PMLR.

\bibitem[{McCorkle and Xygalatas(2013)}]{mccorkle2013mental}
W.W. McCorkle and D.~Xygalatas. 2013.
\newblock \href {https://books.google.ae/books?id=LLfnMgEACAAJ} {\emph{Mental Culture: Classical Social Theory and the Cognitive Science of Religion}}.
\newblock Religion, cognition and culture. Acumen.

\bibitem[{McHugh et~al.(2008)McHugh, Smith, and Sieck}]{mchugh2008cultural}
Anna~P McHugh, Jennifer~L Smith, and Winston~R Sieck. 2008.
\newblock Cultural variations in mental models of collaborative decision making.
\newblock \emph{Macrocognition and naturalistic decision making}, pages 141--158.

\bibitem[{McIntosh et~al.(2024)McIntosh, Liu, Susnjak, Watters, Ng, and Halgamuge}]{10319443}
Timothy~R. McIntosh, Tong Liu, Teo Susnjak, Paul Watters, Alex Ng, and Malka~N. Halgamuge. 2024.
\newblock \href {https://doi.org/10.1109/TAI.2023.3332837} {A culturally sensitive test to evaluate nuanced gpt hallucination}.
\newblock \emph{IEEE Transactions on Artificial Intelligence}, 5(6):2739--2751.

\bibitem[{Minaee et~al.(2024)Minaee, Mikolov, Nikzad, Chenaghlu, Socher, Amatriain, and Gao}]{minaee2024large}
Shervin Minaee, Tomas Mikolov, Narjes Nikzad, Meysam Chenaghlu, Richard Socher, Xavier Amatriain, and Jianfeng Gao. 2024.
\newblock Large language models: A survey.
\newblock \emph{arXiv preprint arXiv:2402.06196}.

\bibitem[{Mishra(2001)}]{mishra2001cognition}
Ramesh~C Mishra. 2001.
\newblock Cognition across cultures.
\newblock \emph{The handbook of culture and psychology}, pages 119--135.

\bibitem[{Moerchen et~al.(2020)Moerchen, Ernst, and Zappella}]{Moerchen2020}
Fabian Moerchen, Patrick Ernst, and Giovanni Zappella. 2020.
\newblock \href {https://www.amazon.science/publications/personalizing-natural-language-understanding-using-multi-armed-bandits-and-implicit-feedback} {Personalizing natural-language understanding using multi-armed bandits and implicit feedback}.
\newblock In \emph{CIKM 2020}.

\bibitem[{Monaghan et~al.(2012)Monaghan, Goodman, and Robinson}]{monaghan2012cultural}
Leila Monaghan, Jane~E Goodman, and Jennifer Robinson. 2012.
\newblock \emph{A cultural approach to interpersonal communication: Essential readings}.
\newblock John Wiley \& Sons.

\bibitem[{Montague et~al.(1970)}]{montague1970universal}
Richard Montague et~al. 1970.
\newblock Universal grammar.
\newblock \emph{1974}, pages 222--46.

\bibitem[{Montalan et~al.(2024)Montalan, Ngui, Leong, Susanto, Rengarajan, Tjhi, and Aji}]{montalan2024kalahihandcraftedgrassrootscultural}
Jann~Railey Montalan, Jian~Gang Ngui, Wei~Qi Leong, Yosephine Susanto, Hamsawardhini Rengarajan, William~Chandra Tjhi, and Alham~Fikri Aji. 2024.
\newblock \href {https://arxiv.org/abs/2409.15380} {Kalahi: A handcrafted, grassroots cultural llm evaluation suite for filipino}.
\newblock \emph{Preprint}, arXiv:2409.15380.

\bibitem[{Mostafazadeh~Davani et~al.(2024)Mostafazadeh~Davani, Diaz, Baker, and Prabhakaran}]{davani2024d3codedisentanglingdisagreementsdata}
Aida Mostafazadeh~Davani, Mark Diaz, Dylan~K Baker, and Vinodkumar Prabhakaran. 2024.
\newblock \href {https://doi.org/10.18653/v1/2024.emnlp-main.1029} {{D}3{CODE}: Disentangling disagreements in data across cultures on offensiveness detection and evaluation}.
\newblock In \emph{Proceedings of the 2024 Conference on Empirical Methods in Natural Language Processing}, pages 18511--18526, Miami, Florida, USA. Association for Computational Linguistics.

\bibitem[{Mukherjee et~al.(2024)Mukherjee, Adilazuarda, Sitaram, Bali, Aji, and Choudhury}]{mukherjee2024culturalconditioningplaceboeffectiveness}
Sagnik Mukherjee, Muhammad~Farid Adilazuarda, Sunayana Sitaram, Kalika Bali, Alham~Fikri Aji, and Monojit Choudhury. 2024.
\newblock \href {https://doi.org/10.18653/v1/2024.emnlp-main.884} {Cultural conditioning or placebo? on the effectiveness of socio-demographic prompting}.
\newblock In \emph{Proceedings of the 2024 Conference on Empirical Methods in Natural Language Processing}, pages 15811--15837, Miami, Florida, USA. Association for Computational Linguistics.

\bibitem[{M{\"u}nch and Smelser(1992)}]{munch1992theory}
Richard M{\"u}nch and Neil~J Smelser. 1992.
\newblock \emph{Theory of culture}.
\newblock University of California Press Berkeley.

\bibitem[{Myung et~al.(2024)Myung, Lee, Zhou, Jin, Putri, Antypas, Borkakoty, Kim, Perez-Almendros, Ayele, Basulto, Ibanez-Garcia, Lee, Muhammad, Park, Rzayev, White, Yimam, Pilehvar, Ousidhoum, Camacho-Collados, and Oh}]{myung2024blendbenchmarkllmseveryday}
Junho Myung, Nayeon Lee, Yi~Zhou, Jiho Jin, Rifki~Afina Putri, Dimosthenis Antypas, Hsuvas Borkakoty, Eunsu Kim, Carla Perez-Almendros, Abinew~Ali Ayele, Victor~Gutierrez Basulto, Yazmin Ibanez-Garcia, Hwaran Lee, Shamsuddeen~Hassan Muhammad, Kiwoong Park, Anar~Sabuhi Rzayev, Nina White, Seid~Muhie Yimam, Mohammad~Taher Pilehvar, Nedjma Ousidhoum, Jose Camacho-Collados, and Alice Oh. 2024.
\newblock \href {https://openreview.net/forum?id=nrEqH502eC} {{BLE}nd: A benchmark for {LLM}s on everyday knowledge in diverse cultures and languages}.
\newblock In \emph{The Thirty-eight Conference on Neural Information Processing Systems Datasets and Benchmarks Track}.

\bibitem[{Nadeem et~al.(2021)Nadeem, Bethke, and Reddy}]{nadeem-etal-2021-stereoset}
Moin Nadeem, Anna Bethke, and Siva Reddy. 2021.
\newblock \href {https://doi.org/10.18653/v1/2021.acl-long.416} {{S}tereo{S}et: Measuring stereotypical bias in pretrained language models}.
\newblock In \emph{Proceedings of the 59th Annual Meeting of the Association for Computational Linguistics and the 11th International Joint Conference on Natural Language Processing (Volume 1: Long Papers)}, pages 5356--5371, Online. Association for Computational Linguistics.

\bibitem[{Nangia et~al.(2020)Nangia, Vania, Bhalerao, and Bowman}]{nangia-etal-2020-crows}
Nikita Nangia, Clara Vania, Rasika Bhalerao, and Samuel~R. Bowman. 2020.
\newblock \href {https://doi.org/10.18653/v1/2020.emnlp-main.154} {{C}row{S}-pairs: A challenge dataset for measuring social biases in masked language models}.
\newblock In \emph{Proceedings of the 2020 Conference on Empirical Methods in Natural Language Processing (EMNLP)}, pages 1953--1967, Online. Association for Computational Linguistics.

\bibitem[{Nisbett and Norenzayan(2002)}]{nisbett2002culture}
Richard~E Nisbett and Ara Norenzayan. 2002.
\newblock Culture and cognition.
\newblock John Wiley \& Sons.

\bibitem[{Noshadi and Dabbagh(2015)}]{noshadi2015metacultural}
Mahdi Noshadi and Ali Dabbagh. 2015.
\newblock Metacultural competence: A benchmark for advances in applied elt.
\newblock In \emph{Selected Papers of 2nd conference on Interdisciplinary Approaches to Language Teaching, Literature and Translation Studies}, pages 52--62. Ferdowsi university of Mashhad, Khate Sefid English Language Group.

\bibitem[{Ortiz et~al.(2013)Ortiz, Beach et~al.}]{ortiz2013ethnographic}
Anna~M Ortiz, Long Beach, et~al. 2013.
\newblock The ethnographic interview.
\newblock In \emph{Research in the college context}, pages 51--64. Routledge.

\bibitem[{Owen et~al.(2024)Owen, Tripathi, Kumar, and Ahmed}]{owen2024komodolinguisticexpeditionindonesias}
Louis Owen, Vishesh Tripathi, Abhay Kumar, and Biddwan Ahmed. 2024.
\newblock \href {https://arxiv.org/abs/2403.09362} {Komodo: A linguistic expedition into indonesia's regional languages}.
\newblock \emph{Preprint}, arXiv:2403.09362.

\bibitem[{Pandey et~al.(2025)Pandey, Budhiraja, Saha, and Choudhury}]{pandey-etal-2025-culturally}
Saurabh~Kumar Pandey, Harshit Budhiraja, Sougata Saha, and Monojit Choudhury. 2025.
\newblock \href {https://aclanthology.org/2025.coling-demos.21/} {{CULTURALLY} {YOURS}: A reading assistant for cross-cultural content}.
\newblock In \emph{Proceedings of the 31st International Conference on Computational Linguistics: System Demonstrations}, pages 208--216, Abu Dhabi, UAE. Association for Computational Linguistics.

\bibitem[{Parsons(1972)}]{parsons1972culture}
Talcott Parsons. 1972.
\newblock Culture and social system revisited.
\newblock \emph{Social Science Quarterly}, pages 253--266.

\bibitem[{Perniss(2018)}]{perniss2018we}
Pamela Perniss. 2018.
\newblock Why we should study multimodal language.
\newblock \emph{Frontiers in psychology}, 9:1109.

\bibitem[{Petroni et~al.(2019)Petroni, Rockt{\"a}schel, Riedel, Lewis, Bakhtin, Wu, and Miller}]{petroni2019language}
Fabio Petroni, Tim Rockt{\"a}schel, Sebastian Riedel, Patrick Lewis, Anton Bakhtin, Yuxiang Wu, and Alexander Miller. 2019.
\newblock \href {https://doi.org/10.18653/v1/D19-1250} {Language models as knowledge bases?}
\newblock In \emph{Proceedings of the 2019 Conference on Empirical Methods in Natural Language Processing and the 9th International Joint Conference on Natural Language Processing (EMNLP-IJCNLP)}, pages 2463--2473, Hong Kong, China. Association for Computational Linguistics.

\bibitem[{Prakken(2006)}]{prakken2006formal}
Henry Prakken. 2006.
\newblock Formal systems for persuasion dialogue.
\newblock \emph{The knowledge engineering review}, 21(2):163--188.

\bibitem[{Pryzant et~al.(2023)Pryzant, Iter, Li, Lee, Zhu, and Zeng}]{pryzant2023automaticpromptoptimizationgradient}
Reid Pryzant, Dan Iter, Jerry Li, Yin Lee, Chenguang Zhu, and Michael Zeng. 2023.
\newblock \href {https://doi.org/10.18653/v1/2023.emnlp-main.494} {Automatic prompt optimization with {\textquotedblleft}gradient descent{\textquotedblright} and beam search}.
\newblock In \emph{Proceedings of the 2023 Conference on Empirical Methods in Natural Language Processing}, pages 7957--7968, Singapore. Association for Computational Linguistics.

\bibitem[{Putri et~al.(2024)Putri, Haznitrama, Adhista, and Oh}]{putri2024llmgenerateculturallyrelevant}
Rifki~Afina Putri, Faiz~Ghifari Haznitrama, Dea Adhista, and Alice Oh. 2024.
\newblock \href {https://doi.org/10.18653/v1/2024.emnlp-main.1145} {Can {LLM} generate culturally relevant commonsense {QA} data? case study in {I}ndonesian and {S}undanese}.
\newblock In \emph{Proceedings of the 2024 Conference on Empirical Methods in Natural Language Processing}, pages 20571--20590, Miami, Florida, USA. Association for Computational Linguistics.

\bibitem[{Rao et~al.(2024)Rao, Yerukola, Shah, Reinecke, and Sap}]{rao2024normadbenchmarkmeasuringcultural}
Abhinav Rao, Akhila Yerukola, Vishwa Shah, Katharina Reinecke, and Maarten Sap. 2024.
\newblock \href {https://arxiv.org/abs/2404.12464} {Normad: A benchmark for measuring the cultural adaptability of large language models}.
\newblock \emph{Preprint}, arXiv:2404.12464.

\bibitem[{Rao et~al.(2023)Rao, Khandelwal, Tanmay, Agarwal, and Choudhury}]{rao-etal-2023-ethical}
Abhinav~Sukumar Rao, Aditi Khandelwal, Kumar Tanmay, Utkarsh Agarwal, and Monojit Choudhury. 2023.
\newblock \href {https://doi.org/10.18653/v1/2023.findings-emnlp.892} {Ethical reasoning over moral alignment: A case and framework for in-context ethical policies in {LLM}s}.
\newblock In \emph{Findings of the Association for Computational Linguistics: EMNLP 2023}, pages 13370--13388, Singapore. Association for Computational Linguistics.

\bibitem[{Rapp et~al.(2021)Rapp, Curti, and Boldi}]{rapp2021human}
Amon Rapp, Lorenzo Curti, and Arianna Boldi. 2021.
\newblock The human side of human-chatbot interaction: A systematic literature review of ten years of research on text-based chatbots.
\newblock \emph{International Journal of Human-Computer Studies}, 151:102630.

\bibitem[{Rawte et~al.(2023)Rawte, Sheth, and Das}]{rawte2023survey}
Vipula Rawte, Amit Sheth, and Amitava Das. 2023.
\newblock \href {https://arxiv.org/abs/2309.05922} {A survey of hallucination in large foundation models}.
\newblock \emph{Preprint}, arXiv:2309.05922.

\bibitem[{Roberts et~al.(2020)Roberts, Raffel, and Shazeer}]{roberts2020much}
Adam Roberts, Colin Raffel, and Noam Shazeer. 2020.
\newblock \href {https://doi.org/10.18653/v1/2020.emnlp-main.437} {How much knowledge can you pack into the parameters of a language model?}
\newblock In \emph{Proceedings of the 2020 Conference on Empirical Methods in Natural Language Processing (EMNLP)}, pages 5418--5426, Online. Association for Computational Linguistics.

\bibitem[{Saha(2024)}]{saha2024persuasive}
Sougata Saha. 2024.
\newblock \emph{Persuasive Dialogue Systems for Social Good}.
\newblock Ph.D. thesis, State University of New York at Buffalo.

\bibitem[{Schein(1990)}]{schein1990organizational}
Edgar~H Schein. 1990.
\newblock \emph{Organizational culture.}, volume~45.
\newblock American Psychological Association.

\bibitem[{Sclar et~al.(2024)Sclar, Choi, Tsvetkov, and Suhr}]{sclar2024quantifyinglanguagemodelssensitivity}
Melanie Sclar, Yejin Choi, Yulia Tsvetkov, and Alane Suhr. 2024.
\newblock \href {https://openreview.net/forum?id=RIu5lyNXjT} {Quantifying language models' sensitivity to spurious features in prompt design or: How i learned to start worrying about prompt formatting}.
\newblock In \emph{The Twelfth International Conference on Learning Representations}.

\bibitem[{Seth et~al.(2024)Seth, Ahuja, Bali, and Sitaram}]{seth2024dosadatasetsocialartifacts}
Agrima Seth, Sanchit Ahuja, Kalika Bali, and Sunayana Sitaram. 2024.
\newblock \href {https://aclanthology.org/2024.lrec-main.474/} {{DOSA}: A dataset of social artifacts from different {I}ndian geographical subcultures}.
\newblock In \emph{Proceedings of the 2024 Joint International Conference on Computational Linguistics, Language Resources and Evaluation (LREC-COLING 2024)}, pages 5323--5337, Torino, Italia. ELRA and ICCL.

\bibitem[{Sewell(2004)}]{sewell2004concept}
William~H Sewell. 2004.
\newblock The concept (s) of culture.
\newblock In \emph{Practicing history}, pages 76--95. Routledge.

\bibitem[{Sharifian(2013)}]{sharifian2013globalisation}
Farzad Sharifian. 2013.
\newblock Globalisation and developing metacultural competence in learning english as an international language.
\newblock \emph{Multilingual education}, 3:1--11.

\bibitem[{Skinner et~al.(2013)}]{skinner2013interview}
Jonathan Skinner et~al. 2013.
\newblock \emph{The interview: An ethnographic approach}, volume~49.
\newblock A\&C Black.

\bibitem[{Slivkins(2019)}]{MAL-068}
Aleksandrs Slivkins. 2019.
\newblock \href {https://doi.org/10.1561/2200000068} {Introduction to multi-armed bandits}.
\newblock \emph{Foundations and Trends® in Machine Learning}, 12(1-2):1--286.

\bibitem[{Soper et~al.(2022)Soper, Pacquetet, Saha, Das, and Srihari}]{soper-etal-2022-lets}
Elizabeth Soper, Erin Pacquetet, Sougata Saha, Souvik Das, and Rohini Srihari. 2022.
\newblock \href {https://doi.org/10.18653/v1/2022.hcinlp-1.5} {Let{'}s chat: Understanding user expectations in socialbot interactions}.
\newblock In \emph{Proceedings of the Second Workshop on Bridging Human--Computer Interaction and Natural Language Processing}, pages 34--39, Seattle, Washington. Association for Computational Linguistics.

\bibitem[{Spencer-Oatey and Franklin(2012)}]{spencer2012culture}
Helen Spencer-Oatey and Peter Franklin. 2012.
\newblock What is culture.
\newblock \emph{A compilation of quotations. GlobalPAD Core Concepts}, 1(22):1--21.

\bibitem[{Spradley(2016)}]{spradley2016ethnographic}
James~P Spradley. 2016.
\newblock \emph{The ethnographic interview}.
\newblock Waveland Press.

\bibitem[{Strauss(1974)}]{strauss1974structural}
C~Levi Strauss. 1974.
\newblock Structural anthropology.
\newblock \emph{Persona \& Derecho}, 1:571.

\bibitem[{Sun et~al.(2020)Sun, Ho, and Lee}]{sun2019lamol}
Fan-Keng Sun, Cheng-Hao Ho, and Hung-Yi Lee. 2020.
\newblock \href {https://openreview.net/forum?id=Skgxcn4YDS} {{\{}LAMAL{\}}: {\{}LA{\}}nguage modeling is all you need for lifelong language learning}.
\newblock In \emph{International Conference on Learning Representations}.

\bibitem[{Sun et~al.(2021)Sun, Ahn, Park, Tsvetkov, and Mortensen}]{sun2020cross}
Jimin Sun, Hwijeen Ahn, Chan~Young Park, Yulia Tsvetkov, and David~R. Mortensen. 2021.
\newblock \href {https://doi.org/10.18653/v1/2021.eacl-main.204} {Cross-cultural similarity features for cross-lingual transfer learning of pragmatically motivated tasks}.
\newblock In \emph{Proceedings of the 16th Conference of the European Chapter of the Association for Computational Linguistics: Main Volume}, pages 2403--2414, Online. Association for Computational Linguistics.

\bibitem[{Tagliamonte(2006)}]{tagliamonte2006analysing}
Sali~A Tagliamonte. 2006.
\newblock \emph{Analysing sociolinguistic variation}.
\newblock Cambridge University Press.

\bibitem[{Tanmay et~al.(2023)Tanmay, Khandelwal, Agarwal, and Choudhury}]{tanmay2023probingmoraldevelopmentlarge}
Kumar Tanmay, Aditi Khandelwal, Utkarsh Agarwal, and Monojit Choudhury. 2023.
\newblock \href {https://arxiv.org/abs/2309.13356} {Probing the moral development of large language models through defining issues test}.
\newblock \emph{Preprint}, arXiv:2309.13356.

\bibitem[{Taylor(2001)}]{taylor2001ethnographic}
Stephanie Taylor. 2001.
\newblock \emph{Ethnographic research: A reader}.
\newblock Sage.

\bibitem[{Thompson et~al.(2020)Thompson, Roberts, and Lupyan}]{thompson2020cultural}
B~Thompson, SG~Roberts, and G~Lupyan. 2020.
\newblock Cultural influences on word meanings revealed through large-scale semantic alignment. nature human behaviour, 4 (10), 1029--1038.

\bibitem[{Thompson et~al.(2011)}]{thompson2011cross}
William~Forde Thompson et~al. 2011.
\newblock Cross-cultural similarities and differences.

\bibitem[{Urban(2010)}]{urban2010method}
Greg Urban. 2010.
\newblock A method for measuring the motion of culture.
\newblock \emph{American Anthropologist}, 112(1):122--139.

\bibitem[{Vaicenavicius et~al.(2019)Vaicenavicius, Widmann, Andersson, Lindsten, Roll, and Sch{\"o}n}]{vaicenavicius2019evaluating}
Juozas Vaicenavicius, David Widmann, Carl Andersson, Fredrik Lindsten, Jacob Roll, and Thomas Sch{\"o}n. 2019.
\newblock Evaluating model calibration in classification.
\newblock In \emph{The 22nd international conference on artificial intelligence and statistics}, pages 3459--3467. PMLR.

\bibitem[{Vanschoren(2019)}]{vanschoren2019meta}
Joaquin Vanschoren. 2019.
\newblock Meta-learning.
\newblock \emph{Automated machine learning: methods, systems, challenges}, pages 35--61.

\bibitem[{Vigliocco et~al.(2014)Vigliocco, Perniss, and Vinson}]{vigliocco2014language}
Gabriella Vigliocco, Pamela Perniss, and David Vinson. 2014.
\newblock Language as a multimodal phenomenon: implications for language learning, processing and evolution.

\bibitem[{Wan et~al.(2023)Wan, Zhao, Chadha, Peng, and Chang}]{wan-etal-2023-personalized}
Yixin Wan, Jieyu Zhao, Aman Chadha, Nanyun Peng, and Kai-Wei Chang. 2023.
\newblock \href {https://doi.org/10.18653/v1/2023.findings-emnlp.648} {Are personalized stochastic parrots more dangerous? evaluating persona biases in dialogue systems}.
\newblock In \emph{Findings of the Association for Computational Linguistics: EMNLP 2023}, pages 9677--9705, Singapore. Association for Computational Linguistics.

\bibitem[{Wang(2021)}]{wang2021meta}
Jane~X Wang. 2021.
\newblock Meta-learning in natural and artificial intelligence.
\newblock \emph{Current Opinion in Behavioral Sciences}, 38:90--95.

\bibitem[{Wang et~al.(2024)Wang, Zhu, Kong, Wei, Yi, Xie, and Sang}]{wang2024cdevalbenchmarkmeasuringcultural}
Yuhang Wang, Yanxu Zhu, Chao Kong, Shuyu Wei, Xiaoyuan Yi, Xing Xie, and Jitao Sang. 2024.
\newblock \href {https://doi.org/10.18653/v1/2024.c3nlp-1.1} {{CDE}val: A benchmark for measuring the cultural dimensions of large language models}.
\newblock In \emph{Proceedings of the 2nd Workshop on Cross-Cultural Considerations in NLP}, pages 1--16, Bangkok, Thailand. Association for Computational Linguistics.

\bibitem[{Welleck et~al.(2024)Welleck, Bertsch, Finlayson, Schoelkopf, Xie, Neubig, Kulikov, and Harchaoui}]{welleck2024decoding}
Sean Welleck, Amanda Bertsch, Matthew Finlayson, Hailey Schoelkopf, Alex Xie, Graham Neubig, Ilia Kulikov, and Zaid Harchaoui. 2024.
\newblock \href {https://openreview.net/forum?id=eskQMcIbMS} {From decoding to meta-generation: Inference-time algorithms for large language models}.
\newblock \emph{Transactions on Machine Learning Research}.
\newblock Survey Certification.

\bibitem[{Whitehead(2005)}]{whitehead2005basic}
Tony~L Whitehead. 2005.
\newblock Basic classical ethnographic research methods.
\newblock \emph{Cultural ecology of health and change}, 1:1--29.

\bibitem[{Wibowo et~al.(2024)Wibowo, Fuadi, Nityasya, Prasojo, and Aji}]{wibowo2024copalidindonesianlanguagereasoning}
Haryo Wibowo, Erland Fuadi, Made Nityasya, Radityo~Eko Prasojo, and Alham Aji. 2024.
\newblock \href {https://doi.org/10.18653/v1/2024.naacl-long.77} {{COPAL}-{ID}: {I}ndonesian language reasoning with local culture and nuances}.
\newblock In \emph{Proceedings of the 2024 Conference of the North American Chapter of the Association for Computational Linguistics: Human Language Technologies (Volume 1: Long Papers)}, pages 1404--1422, Mexico City, Mexico. Association for Computational Linguistics.

\bibitem[{Yang et~al.(2017)Yang, Crain, Berwick, Chomsky, and Bolhuis}]{yang2017growth}
Charles Yang, Stephen Crain, Robert~C Berwick, Noam Chomsky, and Johan~J Bolhuis. 2017.
\newblock The growth of language: Universal grammar, experience, and principles of computation.
\newblock \emph{Neuroscience \& Biobehavioral Reviews}, 81:103--119.

\bibitem[{Yang et~al.(2024)Yang, Lee, Cho, Papailiopoulos, and Lee}]{yang2023predictive}
Seongjun Yang, Gibbeum Lee, Jaewoong Cho, Dimitris Papailiopoulos, and Kangwook Lee. 2024.
\newblock \href {https://openreview.net/forum?id=yUmJ483OB0} {Predictive pipelined decoding: A compute-latency trade-off for exact {LLM} decoding}.
\newblock \emph{Transactions on Machine Learning Research}.

\bibitem[{Yin et~al.(2022)Yin, Bansal, Monajatipoor, Li, and Chang}]{yin2022geomlama}
Da~Yin, Hritik Bansal, Masoud Monajatipoor, Liunian~Harold Li, and Kai-Wei Chang. 2022.
\newblock Geomlama: Geo-diverse commonsense probing on multilingual pre-trained language models.
\newblock In \emph{Proceedings of the 2022 Conference on Empirical Methods in Natural Language Processing}, pages 2039--2055.

\bibitem[{Yin et~al.(2012)Yin, Cui, Li, Yao, and Chen}]{yin2012challenginglongtailrecommendation}
Hongzhi Yin, Bin Cui, Jing Li, Junjie Yao, and Chen Chen. 2012.
\newblock \href {https://doi.org/10.14778/2311906.2311916} {Challenging the long tail recommendation}.
\newblock \emph{Proc. VLDB Endow.}, 5(9):896–907.

\bibitem[{Zhao et~al.(2024)Zhao, Huang, and Li}]{cult_hallu}
Jiajing Zhao, Cheng Huang, and Xian Li. 2024.
\newblock \href {https://doi.org/10.21203/rs.3.rs-4566507/v1} {A comparative study of cultural hallucination in large language models on culturally specific ethical questions}.

\bibitem[{Zheng et~al.(2024)Zheng, Qiu, Shi, and Ma}]{zheng2024towards}
Junhao Zheng, Shengjie Qiu, Chengming Shi, and Qianli Ma. 2024.
\newblock Towards lifelong learning of large language models: A survey.
\newblock \emph{arXiv preprint arXiv:2406.06391}.

\bibitem[{Zhou et~al.(2024)Zhou, Karidi, Garneau, Cao, Liu, Chen, and Hershcovich}]{zhou2024doesmapotofucontain}
Li~Zhou, Taelin Karidi, Nicolas Garneau, Yong Cao, Wanlong Liu, Wenyu Chen, and Daniel Hershcovich. 2024.
\newblock \href {https://arxiv.org/abs/2404.06833} {Does mapo tofu contain coffee? probing llms for food-related cultural knowledge}.
\newblock \emph{Preprint}, arXiv:2404.06833.

\end{thebibliography}

\clearpage
\appendix


\section{Appendix}
\label{sec:appendix}
\noindent
\begin{table*}[b]
\centering
\resizebox{\textwidth}{!}{%
\begin{tabular}{|l|l|l|}
\hline
\textbf{Sl No} & \textbf{Semantic Domain} & \textbf{Full Question}                                                 \\ \hline
1              & Weight Unit              & What is the unit of measuring weight?                                  \\
2              & Drinking Hot Water       & Is it rare or common to see people drink hot water?                    \\
3              & Climate Zone             & Which climate zone does the country belong to?                         \\
4              & Shower Time              & What time of the day people usually take the shower?                   \\
5              & Driving Side             & Which side do people usually keep when driving?                        \\
6              & Household Servants       & Is it rare or common for households to have servants?                  \\
7  & Past Transportation & What was the most popular mode of transportation in the big cities  30 years back? \\
8              & Food Sharing             & Is it rare or common for people to share their food when they eat out? \\
9              & Date Format (Year)       & Does the year appear before/after the month in the date format?        \\
10             & Driver Side              & Which side of the car is the driver seat?                              \\
11             & Broom for Cleaning       & Is it rare or common for people to use broom to clean the floor?       \\
12             & Date Format (Month)      & Does the month appear before/after the year in the date format?        \\
13             & Living with Parents      & Is it rare or common for adults to live with their parents?            \\
14             & Drying Clothes           & How do people dry their wet clothes?                                   \\
15             & Wedding Duration         & Does a wedding ceremony last for more than one day?                    \\
16             & Popular Sports           & What are the most popular sports?                                      \\
17             & Stock Drop Color         & Which color represents the drop in the stock prices?                   \\
18             & Eating Tools             & Which tools do people usually eat food with?                           \\
19             & Height Unit              & What is the unit of measuring height?                                  \\
20             & Meal Tips                & Is it rare or common that customers pay tips after a meal?             \\
21             & Temperature Unit         & What is the unit of measuring temperature?                             \\
22             & Stock Rise Color         & Which color represents the rise in the stock prices?                   \\
23             & Bridal Outfit Color      & What is the color of the bridal outfit in a wedding?                   \\
24 & Funeral Dress       & What is the color of the dress that people wear in a traditional funerals?         \\
25             & Staple Food              & What is the staple food?                                               \\ \hline
\end{tabular}%
}
\caption{Full text of abbreviated questions in Figure \ref{fig:llama_va}.}
\label{tab:domain-question}
\end{table*}



\begin{figure}[!h]
    \centering 
    \includegraphics[width=\columnwidth]{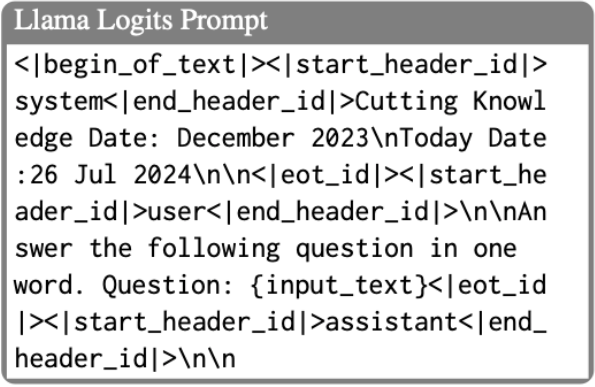} 
    \caption{Prompt used to get GeoMLAMA question logits from Llama.
    }
    \label{fig:llama_promptv2}
\end{figure}

\end{document}